\documentclass[manuscript]{aa}  
\usepackage{graphicx}
\usepackage{subfig}
\usepackage{txfonts}
\usepackage{lineno} 

\usepackage{soul,xcolor}
\usepackage{xcolor}
\usepackage{lineno}
\usepackage{ulem}
\usepackage{natbib}
\bibpunct{(}{)}{;}{a}{}{,} %
\usepackage{hyperref}

\begin{document}

\title{Whistler wave occurrence and the interaction with strahl electrons during the first encounter of Parker Solar Probe}

\author{V.K. Jagarlamudi\inst{1,2}
\and T. Dudok de Wit\inst{1}
\and C. Froment\inst{1}
\and V. Krasnoselskikh \inst{1}
\and A. Larosa\inst{1}
\and L. Bercic \inst{3,4}
\and O. Agapitov\inst{5}
\and J.S. Halekas \inst{6}
\and M. Kretzschmar\inst{1}
\and D. Malaspina\inst{7,8}
\and M. Moncuquet\inst{3}
\and S. D. Bale\inst{5,9,10,11}
\and A. W. Case\inst{12}
\and J. C. Kasper\inst{12,13,14}
\and K. E. Korreck\inst{12}
\and D. E. Larson\inst{5}
\and M. Pulupa\inst{5}
\and M. L. Stevens\inst{12}
\and P. Whittlesey\inst{5}
          }
          

\institute{LPC2E/CNRS, 3 Avenue de la Recherche Scientifique, 45071 Orl\'eans Cedex 2, France\\
\email{vamsee.jagarlamudi@inaf.it}
\and
National Institute for Astrophysics-Institute for Space Astrophysics and Planetology, Via del Fosso del Cavaliere 100, I-00133 Roma, Italy
\and
LESIA, Observatoire de Paris, Universit\'e PSL, CNRS, Sorbonne Universit\'e, Universit\'e de Paris, 5 place Jules Janssen, 92195 Meudon, France
\and
Physics and Astronomy Department, University of Florence, Via Giovanni Sansone 1, I-50019 Sesto Fiorentino, Italy
\and
Space Sciences Laboratory, University of California, Berkeley, CA 94720-7450, USA
\and
Department of Physics and Astronomy, University of Iowa, Iowa City, IA 52242, USA
\and
Laboratory for Atmospheric and Space Physics, University of Colorado, Boulder, CO 80303, USA
\and
Astrophysical and Planetary Sciences Department, University of Colorado, Boulder, CO 80303, USA
\and
Physics Department, University of California, Berkeley, CA 94720-7300, USA
\and
The Blackett Laboratory, Imperial College London, London, SW7 2AZ, UK
\and
School of Physics and Astronomy, Queen Mary University of London, London E1 4NS, UK
\and
Smithsonian Astrophysical Observatory, Cambridge, MA, 02138, USA
\and
Climate and Space Sciences and Engineering, University of Michigan, Ann Arbor, MI 48109, USA
\and
BWX Technologies, Inc., Washington, DC 20002, USA}

\date{Received October 30, 2020; Accepted January 12, 2021}

  \abstract
  {}
   {We studied the properties and occurrence of narrowband whistler waves and their interaction with strahl electrons observed between 0.17 and 0.26 au during the first encounter of Parker Solar Probe.}
   {We used Digital Fields Board (DFB) band-pass filtered (BPF) data from FIELDS to detect the signatures of whistler waves. Additionally parameters derived from the particle distribution functions measured by the Solar Wind Electrons Alphas and Protons (SWEAP) instrument suite were used to investigate the plasma properties, and FIELDS suite measurements were used to investigate the electromagnetic (EM) fields properties corresponding to the observed whistler signatures.} 
   {We observe that the occurrence of whistler waves is low, nearly $\sim 1.5\%$ and less than $0.5\%$ in the analyzed peak and average BPF data, respectively. Whistlers occur highly intermittently and 80$\%$ of the whistlers appear continuously for less than 3 s. The spacecraft frequencies of the analyzed waves are less than 0.2 electron cyclotron frequency ($f_{ce}$
   ). The occurrence rate of whistler waves was found to be anticorrelated with the solar wind bulk velocity. The study of the duration of the whistler intervals revealed an anticorrelation between the duration and the solar wind velocity, as well as between the duration and the normalized amplitude of magnetic field variations. The pitch-angle widths (PAWs) of the field-aligned electron population referred to as the strahl are broader by at least 12 degrees during the presence of large amplitude narrowband whistler waves. This observation points toward an EM wave electron interaction, resulting in pitch-angle scattering. PAWs of strahl electrons corresponding to the short duration whistlers are higher compared to the long duration whistlers, indicating short duration whistlers scatter the strahl electrons better than the long duration ones. Parallel cuts through the strahl electron velocity distribution function (VDF) observed during the whistler intervals appear to depart from the Maxwellian shape typically found in the near-Sun strahl VDFs. The relative decrease in the parallel electron temperature and the increase in PAW for the electrons in the strahl energy range suggests that the interaction with whistler waves results in a transfer of electron momentum from the parallel to the perpendicular direction.}
   {}
\titlerunning{Whistler wave occurrence and the interaction with strahl electrons}
\authorrunning{Jagarlamudi et al.}
\maketitle

\section{Introduction}
Whistler waves are right-handed polarized electromagnetic modes observed between the lower hybrid frequency ($f_{LH}$) and electron cyclotron frequency ($f_{ce}$) in the plasma frame. The range between $f_{LH}$ and $f_{ce}$ is usually referred to as the whistler range since whistler waves are the dominant electromagnetic modes observed in this range. In the solar wind, whistlers are dominantly observed in the range between $f_{LH}$ and $0.5 f_{ce}$ \citep{Zhang1998,Lacombe2014,Tong2019stasticalstudy,Jagarlamudi2020}. 

Whistler wave modes through their interaction with electrons are thought to be one of the prime contributors in the regulation of fundamental processes in the solar wind \citep{Vocks2003,Pagel2007,Kajdic2016,Tang2020}. Wave particle interactions, such as those between whistler waves and electrons, might play a significant role in explaining many physical process such as the heating, acceleration, and scattering of the particles in the solar wind. The electron velocity distribution can be mainly divided into three parts: a low energy isotropic distribution called the core, a high energy isotropic part called the halo, and the heliospheric magnetic field aligned high energy component called the strahl \citep{Feldman1978,Pillip1987a,Pilipp1987b}. While collisions were shown to isotropise the dense low energy electron population referred to as the electron core, they are not sufficient to regulate the more tenuous higher energy electron populations such as the halo and strahl \citep{Ogilvie1978JGR,Pillip1987a}. Wave particle interactions have a crucial role in explaining the phenomena happening in the high energy ranges.

Due to their small mass, electrons reach high thermal velocities in the hot solar corona. The fastest electrons such as strahl can escape the solar corona almost undisturbed, carrying the majority of the heat flux stored in the solar wind. The rate of radial decrease in electron heat flux from the Sun suggests the existence of the scattering mechanism during solar wind expansion \citep{Scime1994JGR,Hammond1996,Stverak2015}. Therefore, understanding the evolution of strahl electrons and the wave modes interacting with them gives us valuable insight into the global solar wind thermodynamics and energy transport. Observations have shown that the strahl pitch-angle width (PAW) increases as we move further from the Sun \citep{Hammond1996,Graham2017,Laura2019}. It is also seen that the relative density of the electron halo increases, while the relative density of the strahl decreases as we move away from the Sun \citep{Maksimovic2005,Stverak2009}. Whistler waves with their interaction with strahl electrons could be able to explain the observed evolution of electron velocity distributions \citep{Vocks2012,Kajdic2016,Boldyrev2019,Tang2020}.

There are quite a few studies on the whistler waves in the free solar wind at 1 au. One of the early studies to show the clear presence of whistler waves in the free solar wind was done by \citet{Zhang1998}. In their study, the authors used the high-resolution electric and magnetic field wave form data on board the Geotail spacecraft. They observed that whistler waves have frequencies between $0.1 f_{ce} $ and $0.4 f_{ce}$, and the wave vectors were dominantly aligned to the magnetic field direction and propagating in the anti-sunward direction. Whistler wave packets were observed for short durations (less than 1 s).

\citet{Lacombe2014}, using the magnetic spectral matrix
routine measurements of the Cluster/STAFF instrument, studied the long duration whistlers (5-10 min). They have studied 20 events, which were observed in the slow wind with a frequency range between $0.1 f_{ce} $ and $0.5 f_{ce}$. The observed waves were quasi-parallel and narrowband. \citet{Tong2019stasticalstudy} carried out a large statistical study of whistler waves using 3 yr of magnetic field spectral data from the ARTEMIS (Acceleration, Reconnection, Turbulence, and Electrodynamics of the Moon’s Interaction with the Sun) spacecraft. They show that the occurrence of whistler waves was dependent on the electron temperature anisotropy, and the amplitude of whistler waves were typically small, below 0.02 of the background magnetic field.

A statistical study on whistler waves in the solar wind in the inner heliosphere (0.3 to 1 au) was performed by \citet{Jagarlamudi2020}, who used the search coil spectral data to identify the signatures of whistler waves. Their observed whistler waves have frequencies between $0.05 f_{ce} $ and $0.3 f_{ce}$. They show different properties of whistler waves and find that the slower the bulk velocity of the solar wind, the higher the occurrence of whistlers. They show that the occurrence probability of whistler waves is lower as we move closer to the Sun and suggest that whistler occurrence and variations in the halo and core anisotropy as well as the heat flux values were related.

\citet{Cattell2020} studied the large amplitude whistler waves in the solar wind at frequencies of 0.2–0.4 $f_{ce}$ using the STEREO electric and magnetic field waveforms. These waves were often observed in association with the stream interaction regions. Their studies show that the large amplitude and obliquely propagating, less coherent whistlers were able to resonantly interact with electrons over a broad energy range. 

A recent study by \citet{Agapitov2020}, using the Parker Solar Probe's (PSP's) magnetic and electric waveform data, has shown the presence of whistler waves when magnetic field dips were observed around switchback boundaries. The observed waves were quasi-parallel to dominantly oblique, wave normal angles were close to the resonance cone. The observed whistler wave packets have frequencies below $0.1 f_{ce}$.

In this study, we focus on the whistler waves observed in the solar wind in the inner heliosphere between $ 0.17$ and $ 0.26$ au using the PSP's first perihelion data. The PSP mission was launched in August 2018 to study the Sun closer than ever before through in situ measurements of solar wind \citep{fox_solar_2016}. 



We present the plasma properties (mainly the strahl electron properties) corresponding to the observed signatures of whistler waves identified using band-pass filtered (BPF) data. Studies by \citet{Lacombe2014} at 1 au show that any local enhancement (concentration of spectral power) observed in the magnetic field power spectral density in the frequency range of $f_{LH}$ and $0.5 f_{ce}$ always corresponded to a narrowband whistler wave. For our study we assume that any local enhancement of spectral power observed in the frequency range between $f_{LH}$ and $0.5 f_{ce}$ is a whistler wave signature \citep{Zhang1998,Lacombe2014,Jagarlamudi2020}. The advantage of using band-pass data is that we have high resolution continuous measurements, which allows us to analyze wave parameters statistically. However, the drawback is that we only have a single component of data available and we do not have the polarization information, which has been compensated for by using the polarization information from the analysis of low time resolution cross-spectral data measured by the search coil magnetometer. Waveform data are useful to show the presence of whistlers in PSP's data. However, only low frequency whistlers can be seen in the continuous waveform (with the sampling rate of 293 $s^{-1}$ and less), that is to say we can use waveforms for special cases, such as when there are drops in the magnetic field \citep{Agapitov2020}. 

This article is structured as follows. In Section 2 we show the data and the methodology followed to identify the whistlers. In Section 3 we present the basic properties of whistler waves, as well as their occurrence and generation conditions. In Section 4, using the strahl distributions, strahl PAW of electrons and the strahl parallel temperatures, we investigate the interaction between whistlers and strahl electrons. In Section 5 we present the conclusions for our study.

\section{Methods and data for the whistler waves analysis}

\begin{figure}
   \sidecaption
   \includegraphics[width=9cm,height=8cm]{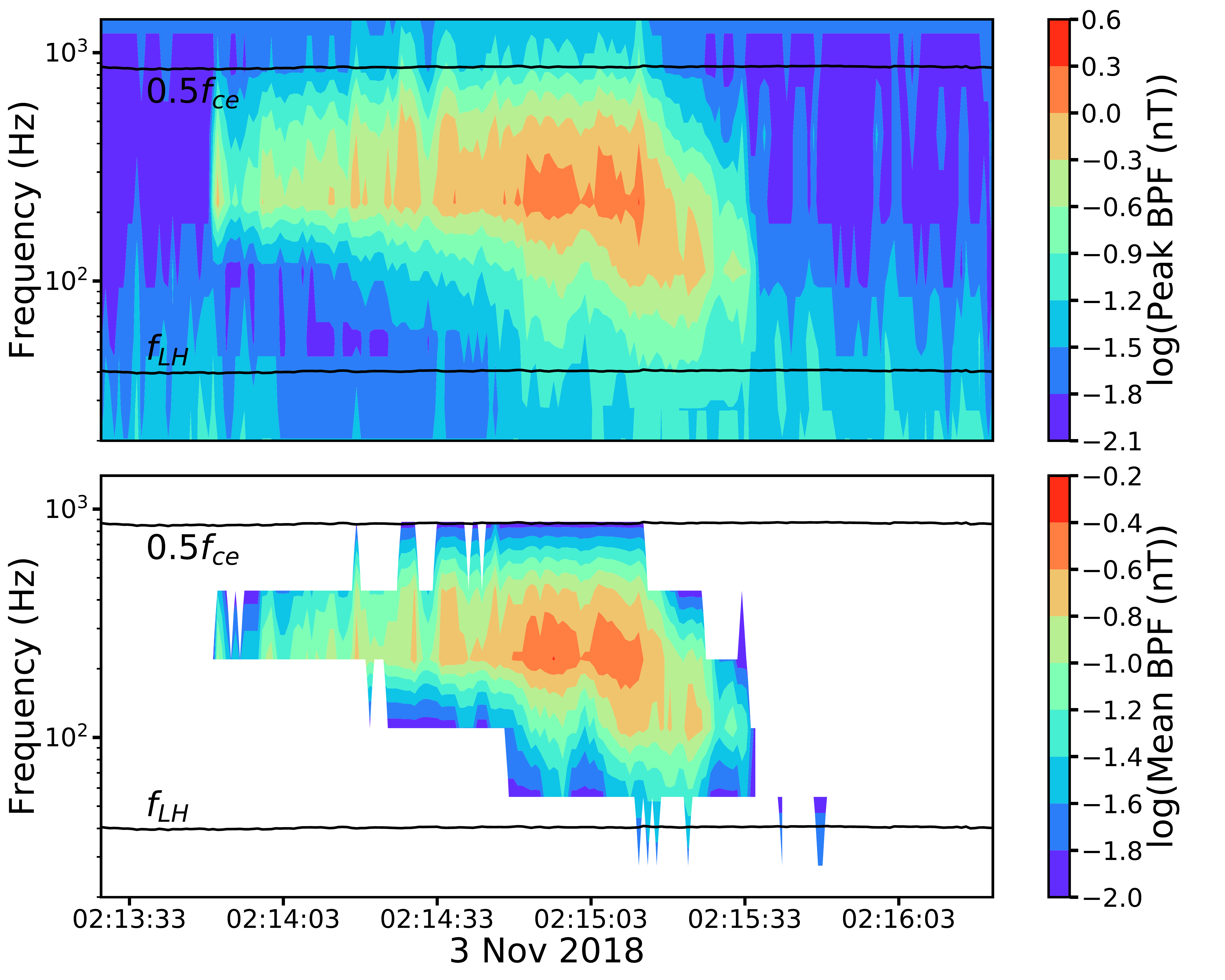}
      \caption{Example of the peak and average band-pass filtered (BPF) data used for the analysis of whistler waves.}
         \label{fig:Mean_peak_data_example}
   \end{figure}

\begin{figure}
   \sidecaption
   \includegraphics[width=9cm]{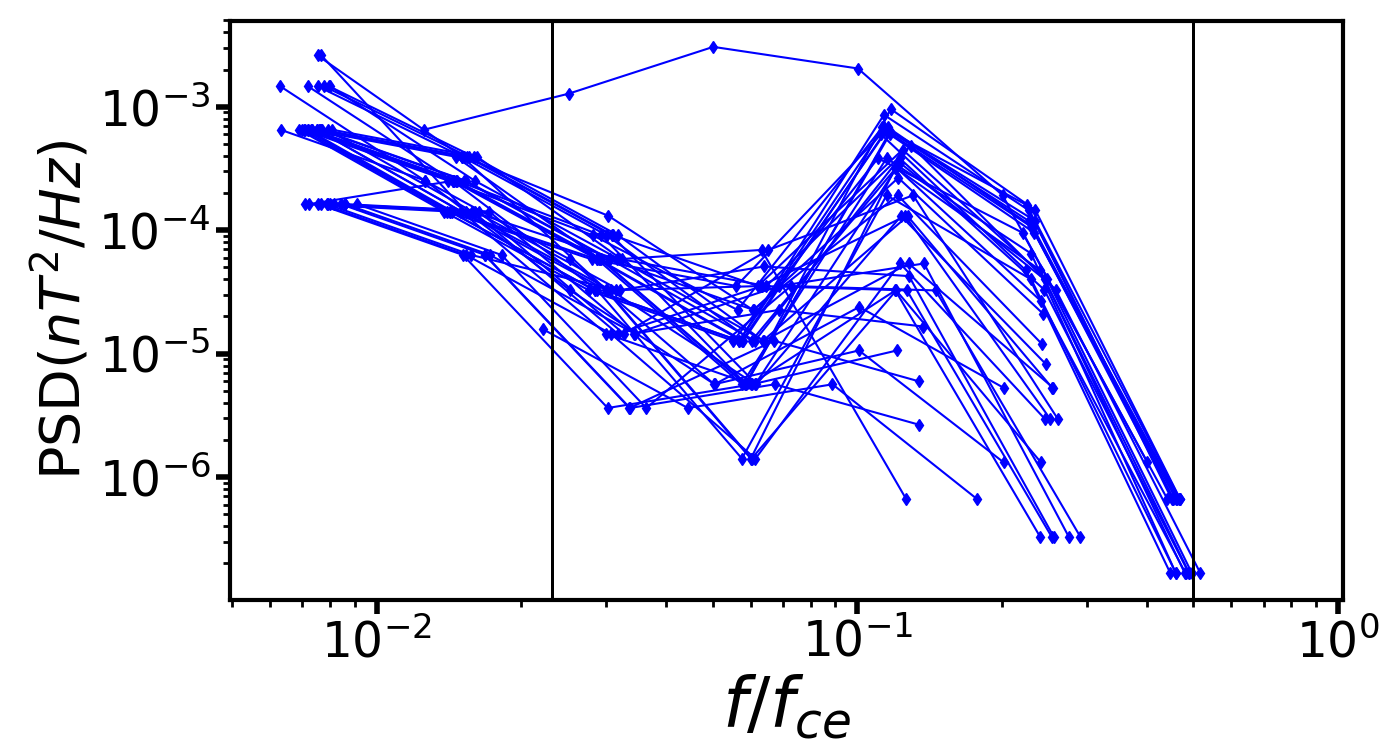}
      \caption{Example of the spectra showing signatures of whistler waves in the average BPF data as a function of $f/f_{ce}$ on November 5, 2018. The black vertical lines correspond to $f_{LH}/f_{ce}$ and 0.5, respectively.
              }
         \label{fig:Whistler_spectra}
   \end{figure}

For the purposes of this paper, whistler analysis was performed using the data from the FIELDS \citep{Bale2016} and SWEAP \citep{Kasper2016} instruments on board PSP during the first encounter with the Sun (October 31 – November 11, 2018). For the detection of the signatures of narrowband whistlers, we used the DC BPF measurements obtained from the Digital Fields Board (DFB) for FIELDS on board PSP \citep{Malaspina2016}.  DFB gives the peak absolute and the average absolute values in each band-pass time series sample of $\sim0.87$ s, covering the frequency range of 0.5 Hz to 9 kHz. The DC BPF data are organized in 15 frequency bins; for each, we have the mean amplitude of the magnetic field and the peak amplitude. BPF data are available for only one component of the magnetic field. Here, we have the Bu component of the SCM \citep{jannet2020}. However, the main advantage compared to the three component cross-spectral data ($\sim 28 $ s) is that BPF data are of a higher time resolution ($\sim 0.87 $ s)  and importantly both the peak and average data are available.

In Figure \ref{fig:Mean_peak_data_example} we show an example of peak and average BPF data using a 3 minute interval when the BPF spectra showed a local enhancement of power. The data output of average BPF data is zero when the signal is very low in the corresponding frequency channels.

For the detection of the whistler signatures, we used a similar method as those used in the studies of \citet{Jagarlamudi2020} and \citet{Tong2019stasticalstudy},  where a local enhancement of spectral power in the whistler range was inferred to indicate the presence of a narrowband whistler wave. First, we squared the peak and average BPF data and divided the squared values by the corresponding frequency bin width, which gives us an equivalent of the power spectral density (PSD) values. Using the PSD values, we identified the presence of one single local maximum in the whistler range which clearly stands out with respect to the PSD of the background turbulence \citep{Alexandrova2012,Alexandrova2020}. Mathematically speaking, the indicator for whistler wave influenced PSD spectra is, as we go toward higher frequencies at a certain frequency, $\frac{dPSD}{df}$ will be positive and then it naturally becomes negative again. However, we would like to mention that the suggested method could only be used with higher confidence for the average spectra. The reason is that the observations to date have shown the presence of whistlers in waveform data to the presence of local enhancement of spectral power in the average spectral data only.

An example of average PSD spectra which contain whistlers is shown in Figure \ref{fig:Whistler_spectra}, where spectra with distinctive local enhancement in the whistler range were selected. After selecting the spectra with whistler signatures, we studied the plasma properties corresponding to those whistler signatures. We mainly focused on the strahl electron properties.

The advantage of BPF data is that thanks to their time resolution of $\sim 0.87$ s, they gives us a much better statistics and information on the duration of the whistlers compared to the low resolution cross-spectral data ($\sim 28$ s). However, the cross-spectral data give us the approximate information on the absolute ellipticity. The study of cross-spectral data from PSP's first perihelion by \citet{froment2020whistler} has shown that all the cross-spectra, which have shown the local enhancement of power in the whistler range, have higher ellipticity ($\sim 1$), indicating circular polarization. This supports our assumption that spectra with local maxima in the BPF data in the whistler range are most probably due to the whistler waves. 

To study the whistler wave properties, we used DC magnetic field data from the fluxgate magnetometer on board PSP \citep{Bale2016}. We used electron and proton observations made by the Solar Wind Electrons Alphas and Protons (SWEAP) experiment \citep{Kasper2016}. Proton densities and proton bulk velocities were obtained from their respective distribution functions measured by the Solar Probe Cup (SPC) at a cadence of $\sim 0.22$ s \citep{case_solar_2020}. The electron velocity distribution functions were measured by Solar Probe Analyzer (SPAN) electron sensors on the ram (ahead) and anti-ram (behind) faces of the spacecraft \citep{whittlesey_solar_2020} and the data available for the first perihelion was of  $\sim28 $ s cadence.

We used the 4 Hz magnetic field data and interpolated these to the resolution of available BPF data. For proton and electron parameters, we considered the closest available value to the BPF interval with the whistler signature. The magnetic field data and the proton moments are taken from \href{https://sppgway.jhuapl.edu/}{PSP Science Gateway}. 

The electron density, core temperatures, and heat flux are taken from the work of \citet{Halekas2020} and \citet{Halekas2020arXiv}, where a bi-Maxwellian distribution is assumed to fit the core parameters.  The strahl pitch-angle widths (PAWs), the cuts through the strahl electron VDFs, and the strahl parallel temperatures ($T_{s\parallel}$) are taken from the work of \citet{Bercic2020}. PAWs represent the full-width-at-half-maximum of a Gaussian fit to pitch-angle distribution functions at every instrument energy bin. The maximal values of these fits by definition appear at a pitch-angle of 0 deg and thus form the parallel cut through the strahl VDF. In using this technique to study the properties of the strahl along the magnetic field, we account for the portion of the strahl VDF which is sometimes blocked by the spacecraft heat shield (see \citet{Bercic2020} for more details about the analysis). We note that $T_{s \parallel}$ is a Maxwellian temperature of the strahl along the magnetic field direction.

For our analysis, we also use the low‐frequency receiver (LFR) data from the Radio Frequency Spectrometer on board PSP \citep{Pulupa2017}. Using the $\sim 7$ s resolution LFR data, we detect the presence of Langmuir waves. The technique followed for the detection of Langmuir waves is similar to the whistler wave detection, that is to say usually the spectra influenced by Langmuir waves appear with a very distinctive local maxima (at least 1 order of magnitude higher than the regular QTN line peak ); using this as an indicator, we looked for large local enhancements in the LFR intervals between 0.9$f_{pe}$ and 1.1$f_{pe}$.

\section{Whistler wave occurrence and their properties}
\subsection{General properties}
We have detected 2492 and 17313 spectra with the whistler signatures in the 1142095 average/peak samples of BPF data. We observed that whistlers occur intermittently. The spectra which showed the whistler signatures represent less than $0.5\%$ of the average BPF data and around $\sim 1.5\%$ of the peak BPF data. The reason for the relatively higher number of whistlers observed in the peak BPF data compared to the average BPF data is understandable, since if the whistlers have a low amplitude or a very short lifetime, they are not visible in the average spectra, but they can only be observed in the peak data. However, if the whistlers are of a large amplitude or long duration, they would appear in average BPF data and in peak BPF data. 

From both the average and peak spectra showing the whistler waves signatures, we observe that the occurrence of whistler waves is low ($<2\%$). This is in line with the predictions by \citet{Jagarlamudi2020}, where the authors suggest that for a distance lower than 0.3 au from the Sun, the occurrence rate should be less than observed $3\%$ at 0.3 au. The reason the authors suggest that is because as we go closer to the Sun, the conditions for whistler generation through WHFI and WTAI are weaker; therefore, the closer we are to the Sun, the lower the occurrence rate of whistlers. All whistler studies to date which used spectral data have used average spectral data. So while comparing our observed properties with other studies, we used the properties of whistlers identified in average BPF data.   

\begin{figure}%
\centering
\subfloat(a){%
\label{fig:first}%
\includegraphics[width=0.8\linewidth]{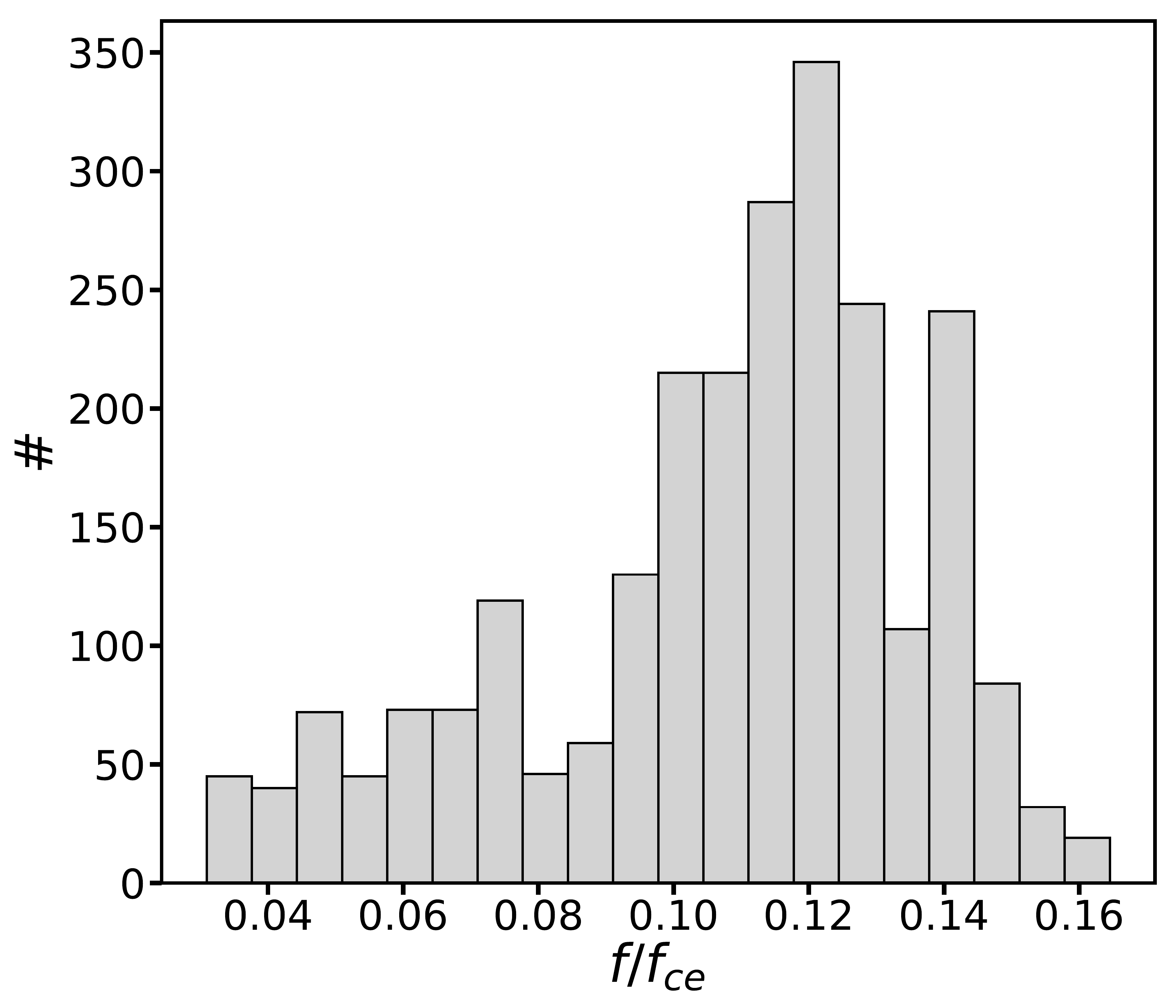}}%
\qquad
\subfloat(b){%
\label{fig:second}%
\includegraphics[width=0.8\linewidth]{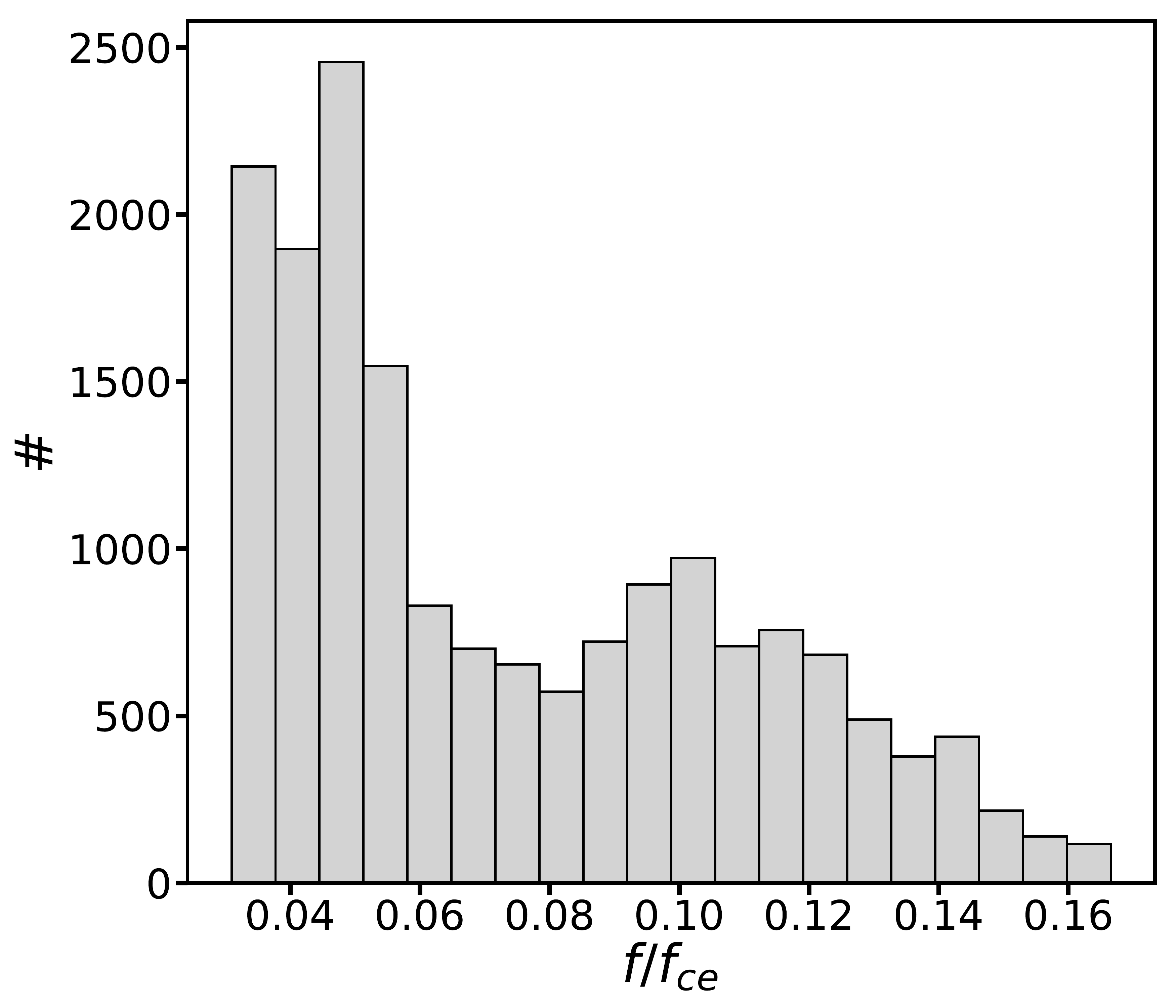}}%
\caption{Histogram of normalized frequencies of the whistler waves in the average and peak BPF data. We show the normalized frequency of whistlers observed in average BPF data in panel (a) and the normalized frequency of whistlers observed in the peak BPF data in panel (b).}
\label{fig:Mean Peak Normalized frequencies}
\end{figure}

In Figure \ref{fig:Mean Peak Normalized frequencies} (a) and (b), we show the normalized spacecraft frequency of the whistler waves identified using the peak and average BPF data. From Figure \ref{fig:Mean Peak Normalized frequencies} (a), we observe that most of the whistlers in the average BPF data are concentrated around 0.1 $f_{ce}$, which is similar to previous observations in the solar wind \citep{Tong2019stasticalstudy,Jagarlamudi2020}. Meanwhile, from Figure \ref{fig:Mean Peak Normalized frequencies} (b), we can observe that peak BPF data show a significant fraction of whistlers in the low normalized frequency ($<0.05$) range.  

\begin{figure}%
\centering
\subfloat(a){%
\label{fig:first}%
\includegraphics[width=0.8\linewidth]{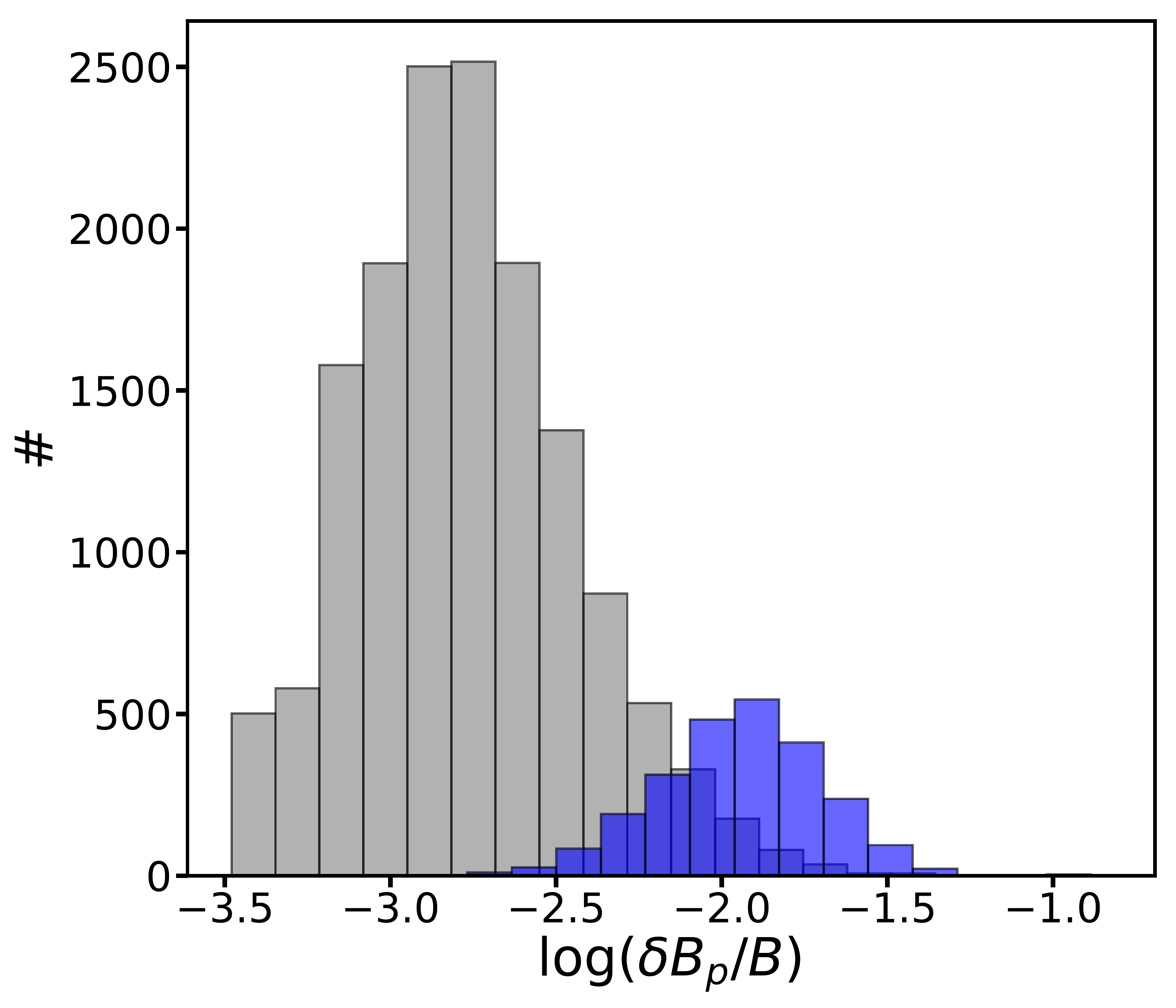}}%
\qquad
\subfloat(b){%
\label{fig:second}%
\includegraphics[width=0.8\linewidth]{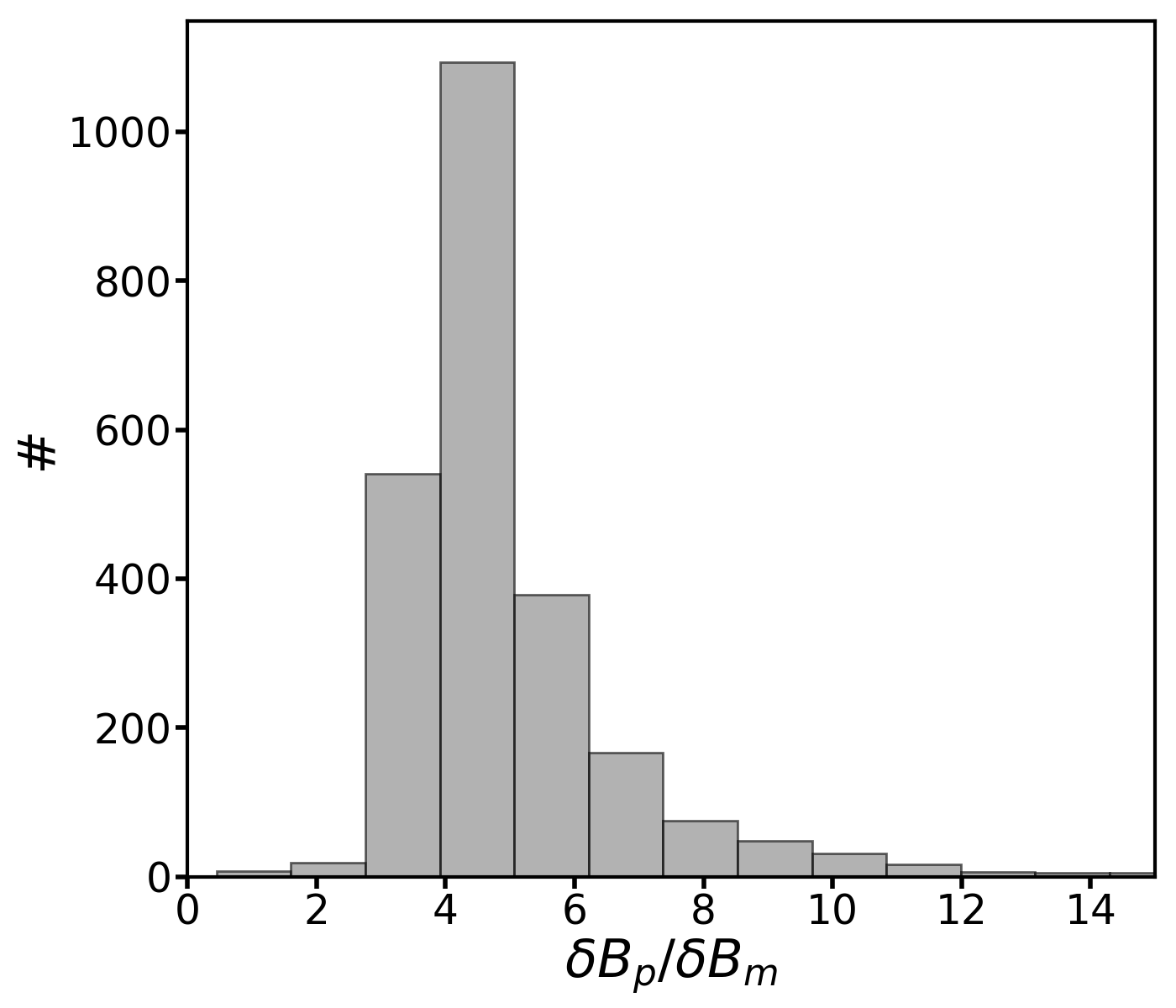}}%
\caption{Histogram of peak and average amplitudes. In panel (a), we show the log of normalized amplitudes of whistlers observed in the peak; blue corresponds to the data where whistlers are observed in both peak and average BPF, and gray corresponds to when the whistlers are observed only in the peak BPF spectra. In panel (b), we show the ratio of the peak and the average amplitude of spectra with whistler signatures.}
\label{fig:Mean peak Histogram amplitudes}
\end{figure}

In Figure \ref{fig:Mean peak Histogram amplitudes} (a) and (b), we show a histogram of the log of normalized peak amplitudes and the ratio of peak and average amplitudes corresponding to the whistlers. The method followed to estimate the approximate amplitude of the fluctuations that are associated with whistler waves is as follows: The spectral value of the identified local maximum is multiplied with its respective frequency of the wave and the square root of this value is interpreted as the amplitude of the fluctuation. We note that $\delta B_p$ represents the peak amplitude calculated using the peak BPF data and $\delta B_m$ represents the average amplitude calculated using the average BPF data.

In Figure \ref{fig:Mean peak Histogram amplitudes} (a), the blue histogram corresponds to intervals when whistlers are observed both in the peak and average BPF data, whereas the gray histogram corresponds to intervals when the whistlers are only observed in the peak BPF data, but not in the average. We observe a clear separation between the two distributions. Normalized peak amplitudes of the whistlers, which are only observed in the peak band-pass data, are smaller (gray) than the ones which are observed in both peak and average BPF data (blue). We observe that most of the whistlers are of a low amplitude, and these whistlers are not observed in the average BPF data. We cannot deduce whether the low-amplitude whistlers are short-lived or not. However, we can observe that there is a considerable overlap between the gray and blue histograms, which suggests that there are whistlers which might be of a large enough amplitude to be visible in average BPF data, but they are very short-lived. Therefore, they are not visible in the average BPF data.  
We can also understand that most of the low normalized frequency whistlers observed in Figure \ref{fig:Mean Peak Normalized frequencies} (b) are of a low amplitude and that is the reason they are not visible in the average BPF data.

In Figure \ref{fig:Mean peak Histogram amplitudes} (b), we show the ratio of peak and average amplitude of the whistlers when whistlers are observed simultaneously in the average and peak BPF data. This relation is important in understanding the variability of the envelope of the whistler
wave. From the plot, we observe that their ratios are concentrated between 3 to 7. This shows that when the whistler signatures are observed in both average and peak data, the ratios are nearly constant and there is no high variability. This leads us to conclude that there might not be high variability in the whistler envelopes in our study when the whistlers are observed in both peak and average BPF data.  

\begin{figure}
   \sidecaption
   \includegraphics[width=8cm]{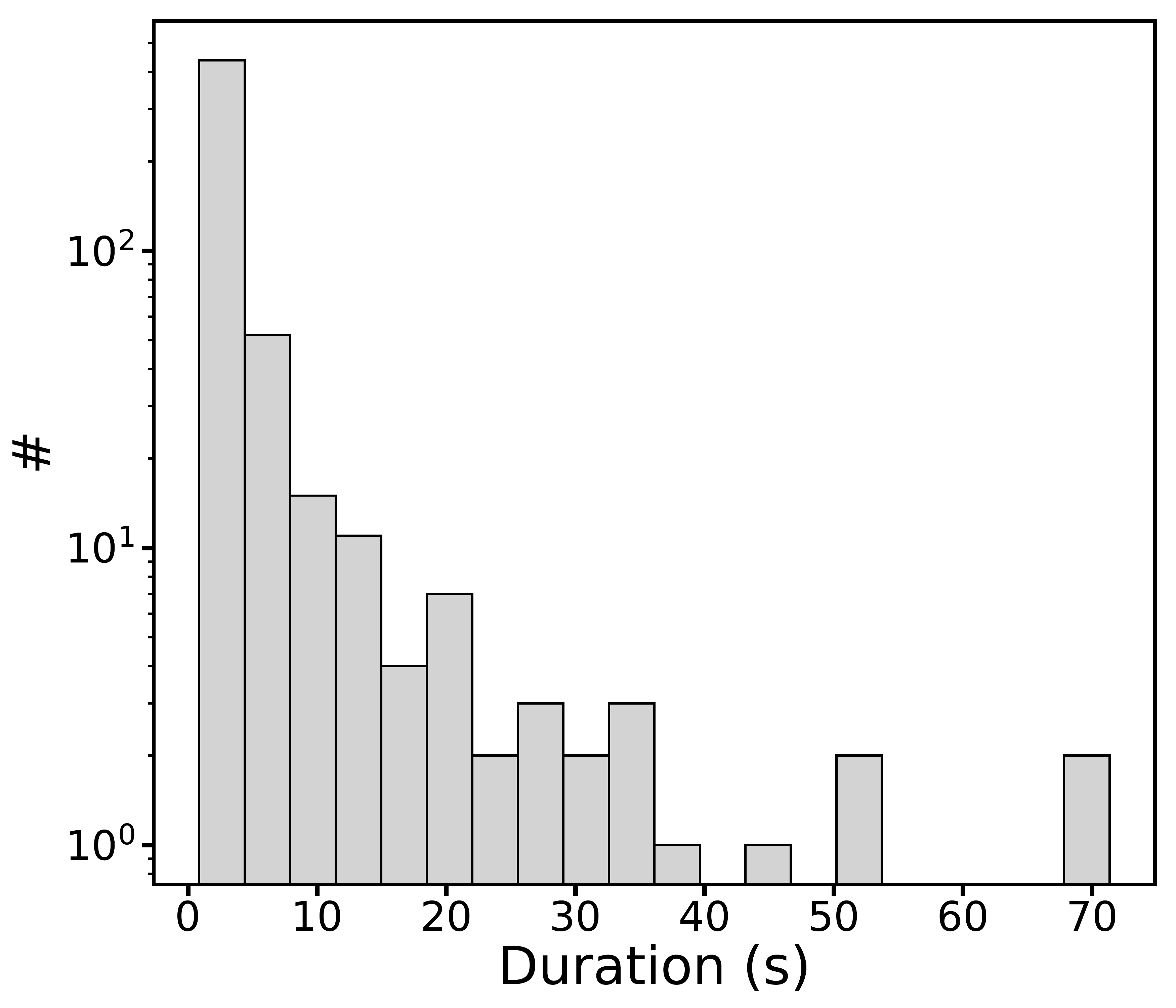}
      \caption{Histogram of the duration of whistler wave's continuous observations in the average BPF data. }
         \label{fig:whistler_lifetime}
   \end{figure}

In Figure \ref{fig:whistler_lifetime} we show a histogram of whistlers as a function of the duration of their observation. The minimum whistler duration is dependent on the resolution of BPF data; therefore, the minimum duration of the whistler is $\sim 0.87$ s. For this study we use the whistlers observed in the average BPF data, as this provides the only approximate representation of how long the whistlers are continuously observed. Most of the whistlers in average BPF data are of a comparatively large amplitude and occur for a time that is long enough to be observed in average BPF spectra. We observe that $80\%$ of the time, whistlers occur continuously for less than 3 s and the probability of observing whistlers continuously for a long duration ($>30$ s) is low. This shows that most of the whistlers occur intermittently, and the probability that whistlers occur for a long duration is low. There is an exponential decrease in the duration of the time whistlers are continuously observed. However, even when the whistlers appear continuously in the BPF data, it does not necessarily mean that a large whistler wave packet is present for such a long period. We believe that it could be a continuous occurrence of short duration whistler wave packets for a long time.

We have presented some of the basic features of the observed whistler signatures. Now we explain how we studied the properties of whistlers as a function of different physical parameters. First, we directly related the presence of whistlers to their observed conditions. Second, we related the plasma conditions to the duration of a consecutive whistler appearance. The first one gives information on the conditions when the whistlers are observed, while the second one gives important information on the differences in the conditions when whistlers are observed for a short duration compared to when they are observed continuously for a long duration. For these studies, we used the whistlers observed in the average BPF data, the reason is that they represent the whole interval size unlike the whistlers that are only observed in the peak BPF.

\begin{figure}
   \sidecaption
   \includegraphics[width=8cm]{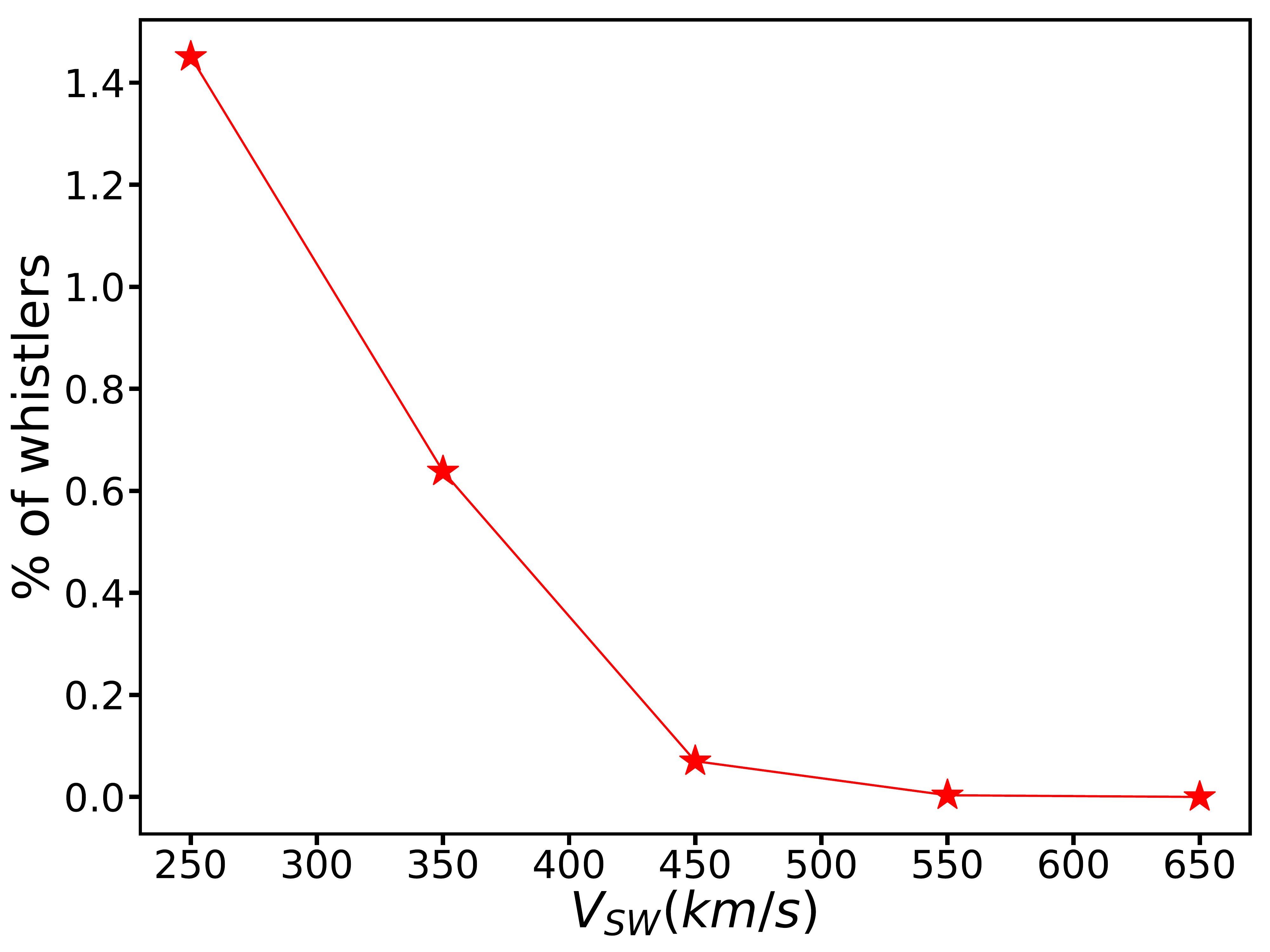}
      \caption{ Occurrence rate of whistler waves as a function of the solar wind bulk speed ($V_{sw}$). For each velocity bin, we show the fraction of BPF data that have the signature of whistler waves. }
         \label{fig:whistler_percetage}
   \end{figure}

In Figure \ref{fig:whistler_percetage} we show the percentage of whistlers as a function of solar wind bulk velocity, that is to say the number of whistlers in the velocity bin to the number of spectra available in the corresponding velocity bin, which takes the occurrence rate of different solar wind speeds into account. We observe that the lower the velocity of the wind, the higher the presence of whistler waves. A similar behavior is shown in the study of \citet{Jagarlamudi2020}, where the authors show the anticorrelation between the occurrence of whistler waves and the solar wind velocity. The authors explain the reason why for slower wind speeds, the conditions were better for the generation of whistlers through whistler temperature anisotropy instability (WTA) and whistler heat flux instability (WHFI).

We also looked into the relation between magnetic field gradients (such as drops, jumps and discontinuities) and the occurrence of whistlers. We used the ratio $\lvert\frac{B-\langle B \rangle}{\langle B \rangle}\rvert$ as an indicator for the magnetic field gradients, and we observe that nearly $\sim 80\%$ of the time whistlers appear, the normalized magnetic field variations ($\lvert\frac{B-\langle B \rangle}{\langle B \rangle}\rvert$) are less than $30 \%$, that is to say most of the whistlers appear when there are not any large absolute magnetic field gradients. This guided us in concluding that large magnetic field gradients are not necessary for the occurrence of whistlers.

We also studied the relation between the occurrence of whistlers and the structures with sudden changes in the radial magnetic field orientation, called switchbacks \citep{bale19, kasper19, DudokdeWit2020}. For this, we used the switchbacks identified in the work of \citet{larosa_switchbacks_2020}. We observed that only $\sim15\%$ of the switchbacks showed the presence of whistler waves close to their boundaries or inside the structure.

\begin{figure}
   \sidecaption
   \includegraphics[width=9cm]{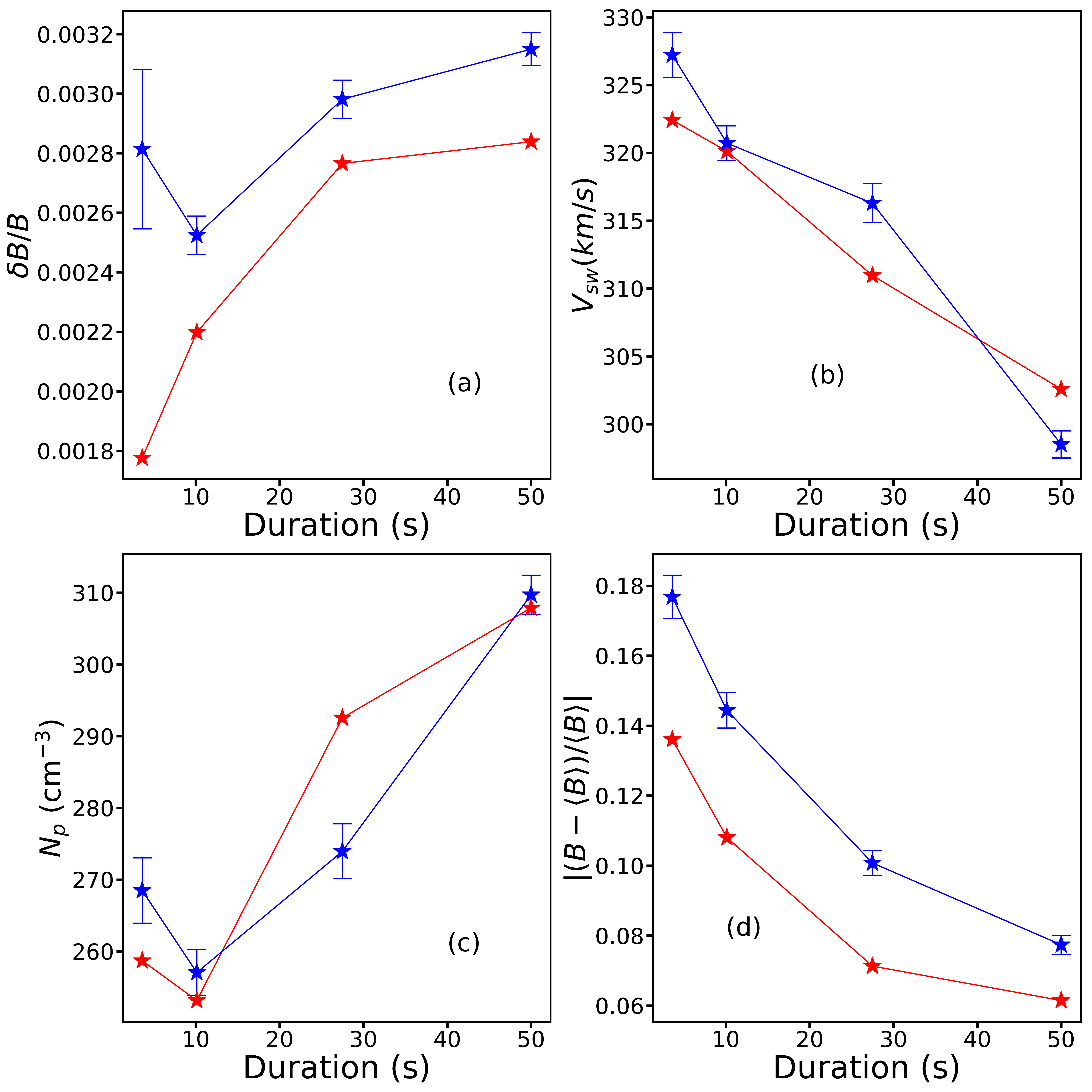}
      \caption{Different properties of whistlers as a function of the duration of their observations; blue and red correspond to the mean and median values, respectively. In panel (a), we show the normalized amplitude of whistler waves; in panel (b), the bulk velocity of the solar wind is shown; in panel (c), the density of the solar wind is shown; and, in panel (d), the magnetic field variations ($\lvert\frac{B-\langle B \rangle}{\langle B \rangle}\lvert$) are shown. Error bars show the standard error ($\frac{\sigma}{\sqrt{n}}$).
              }
         \label{fig:duration_properties}
   \end{figure}

In Figure \ref{fig:duration_properties} we show the mean (blue) and median (red) of different plasma parameters when the whistlers are observed, as a function of their duration. We observe that the normalized amplitudes of the whistler waves which occur continuously are slightly higher compared to the whistlers of short duration. The velocity of the solar wind corresponding to the whistlers is relatively lower for the cases when the whistlers are observed continuously for a long duration. The density is higher for long duration whistlers. We also observe that normalized magnetic field variations ($\lvert\frac{B-\langle B \rangle}{\langle B \rangle}\rvert$) are lower when the whistler waves are observed for a long duration. 
Similarly, we have also studied the variations in the radial magnetic field as a function of the duration of the whistlers (not shown here). We observed that radial magnetic field variations were lower when the whistlers were observed for a long duration. This indicates that the probability of long duration whistlers is lower when there are switchbacks. Long duration whistlers occur when the conditions are quiet. Now, in the following subsection, we look into the possible generation mechanism for the observed whistlers.

\subsection{Whistler wave generation}

\begin{figure}
   \sidecaption
   \includegraphics[width=8cm]{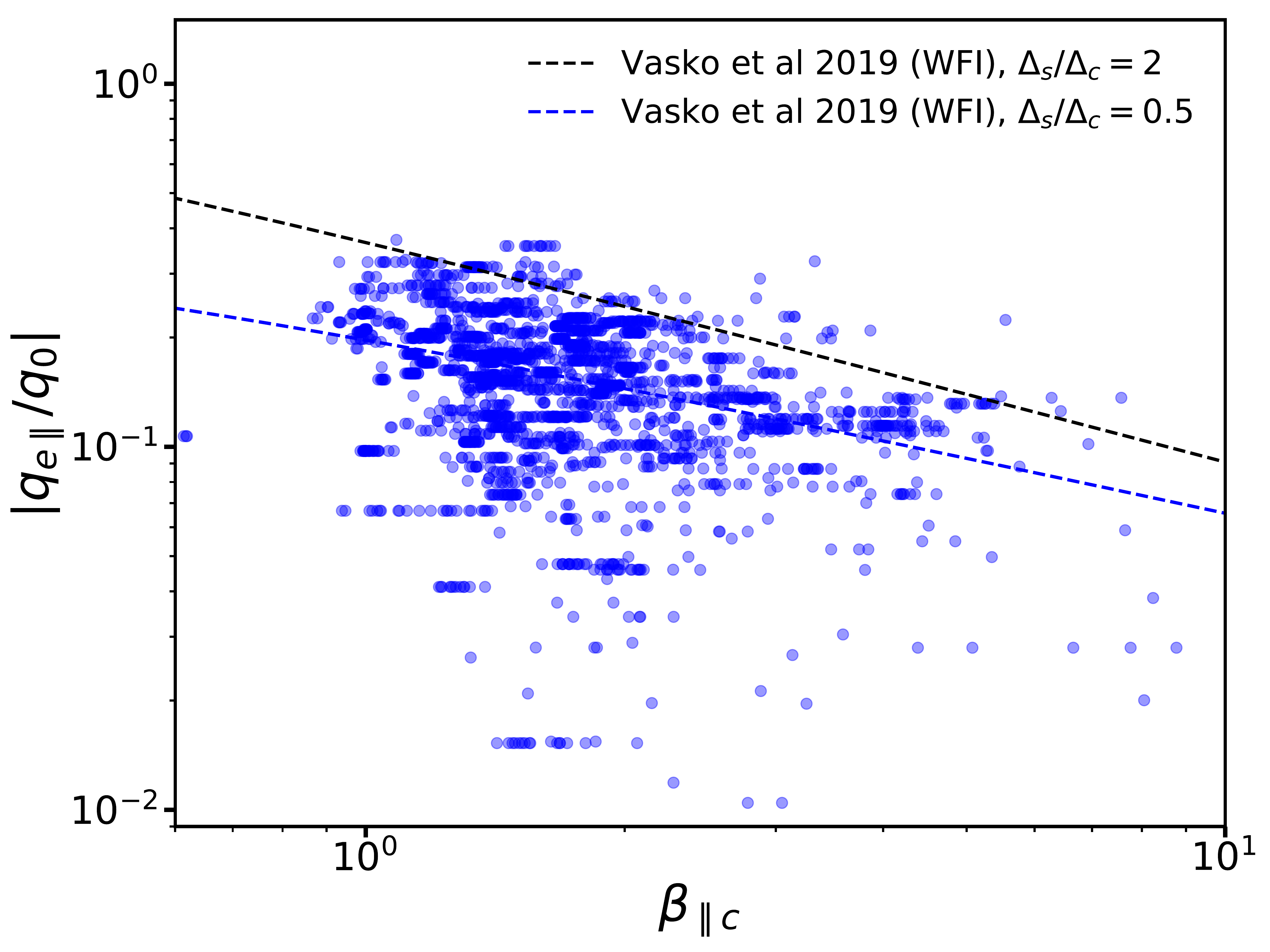}
      \caption{Normalized heat flux of the whistlers observed in the average BPF data. We show the normalized heat flux ($q_{e\parallel}/q_0$) as a function of electron core parallel beta ($\beta_{e\parallel c}$), the dashed lines correspond to the thresholds of whistler fan instability for $\Delta_s/\Delta_c=0.5$ and $\Delta_s/\Delta_c=2$, given by \citet{Vasko2019}. }
         \label{fig:whistler instabilities}
   \end{figure}

Studies by \citet{Lacombe2014}, \citet{Stansby2016}, \citet{Tong2019stasticalstudy,Tong2019} and \citet{Jagarlamudi2020} show that whistler heat flux instability is at work when whistlers are observed. For our case, in which we studied the whistlers which are observed closer to the Sun, we do not have an accurate estimate for the whistler heat flux instability threshold in the literature yet. The level of threshold is sensitive to variations in the densities of electron core and halo populations, and also their temperatures \citep{Gary1994JGR}, which vary with radial distance. Therefore, we could not verify whether the whistler heat flux instability is at work or not. However, using the work of \citet{Vasko2019}, where the instability thresholds were estimated by considering the electron core-strahl velocity distribution functions typical for the solar wind closer to the Sun (0.3- 0.4 au), we could verify the probability of whistler fan instability (for oblique whistlers) working. For our study, we used the normalized heat flux values from the work of \citet{Halekas2020arXiv}. In Figure \ref{fig:whistler instabilities} we present the normalized heat flux as a function of electron core parallel beta ($\beta_{\parallel c}$).

From Figure \ref{fig:whistler instabilities} we infer that the whistler fan instability (for oblique whistlers) is probably at work, as the whistler intervals are around those thresholds \citep{Vasko2019}. However, fan instability is only pronounced for oblique whistlers and we do not have information on the angle of propagation. Therefore, it is important to know the angle of the whistler wave propagation with the mean magnetic field to properly identify for which cases the fan instability is at work. This will be one of the important goals for the future study.

Recent study by \citet{Jagarlamudi2020} have suggested that whistler core and halo anisotropy instabilities might be at work when the whistlers are observed. For our study, only core electron anisotropy values are available and they are of a low resolution; therefore, we are not able to identify whether the whistler anisotropy instabilities are the source of our observed whistlers.

While analyzing the electron parameters, such as the density and temperature measurements obtained from the QTN technique \citep{moncuquet20}, to study the conditions of whistler generation, we observed that for most of the times when the whistlers were observed, no halo or core temperatures were available corresponding to the whistler interval. This frequently observed behavior led us to probe the LFR data used for the electron parameter estimation in the QTN technique. Interestingly, we observed that the LFR spectra showed the presence of a large spectral enhancement around the electron plasma frequency whenever there was a whistler wave during that time period. We identify these enhancements in LFR spectra as Langmuir waves, as all the jumps are centered around an electron plasma frequency (0.9 to 1.1 $f_{pe}$). The simultaneous presence of whistlers and Langmuir waves is similar to what has been reported in the study of \citet{Kennel1980} using the data from ISEE-3. 

\begin{figure}
   \sidecaption
   \includegraphics[width=9cm]{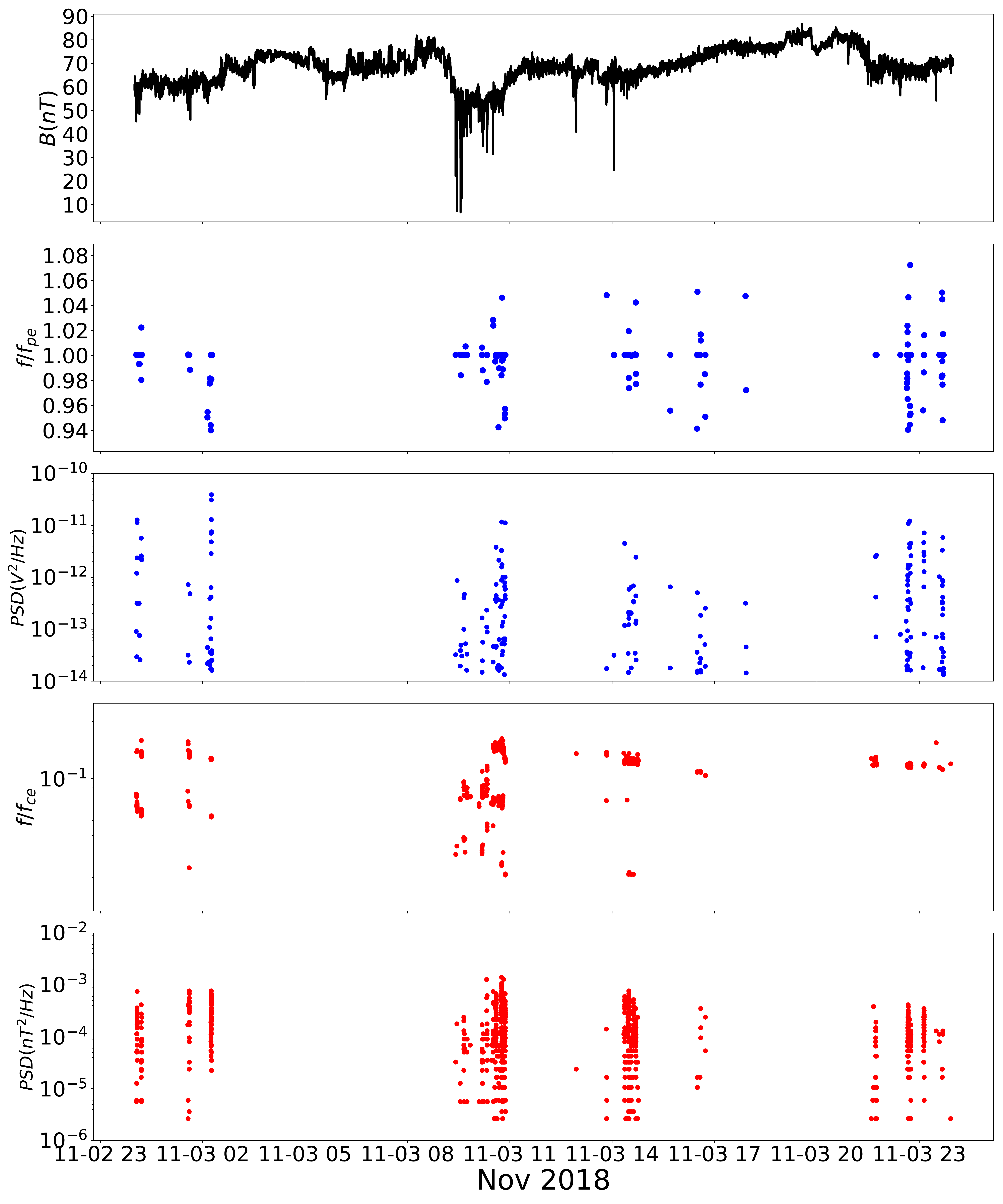}
      \caption{Example of simultaneous observation of whistlers and Langmuir waves in the PSP data. Panel 1 shows the absolute magnetic field, panel 2 shows the normalized frequency of Langmuir waves with electron plasma frequency, panel 3 shows the spectral density of the observed Langmuir waves, panel 4 shows the normalized frequency of whistler waves with the electron cyclotron frequency and panel 5 shows the spectral density of whistler waves.
              }
         \label{Whistler_Langmuir}
   \end{figure}

We identified that 85$\%$ of the time, intervals corresponding to whistlers in the average BPF showed the presence of Langmuir waves.  A glimpse of this behavior can be seen in Figure \ref{Whistler_Langmuir}, where the Langmuir waves normalized frequency and their PSD along with the whistlers normalized frequency and their PSD is shown in blue and red. From this plot, we can infer that there is a clear correlation between the occurrence of whistlers (red dots) and Langmuir waves (blue dots) during this interval. These simultaneous observations of whistlers and Langmuir waves give us a clue that there might be a common generation source for both the whistlers and Langmuir waves; therefore, we have to broaden our ideas as to potential whistler generation sources.  An advanced study will be performed in the future using the waveform and high resolution particle data to accurately identify the source of the whistlers closer to the Sun. 

\section{Whistlers and strahl electrons}

\begin{figure}
   \sidecaption
   \includegraphics[width=9cm]{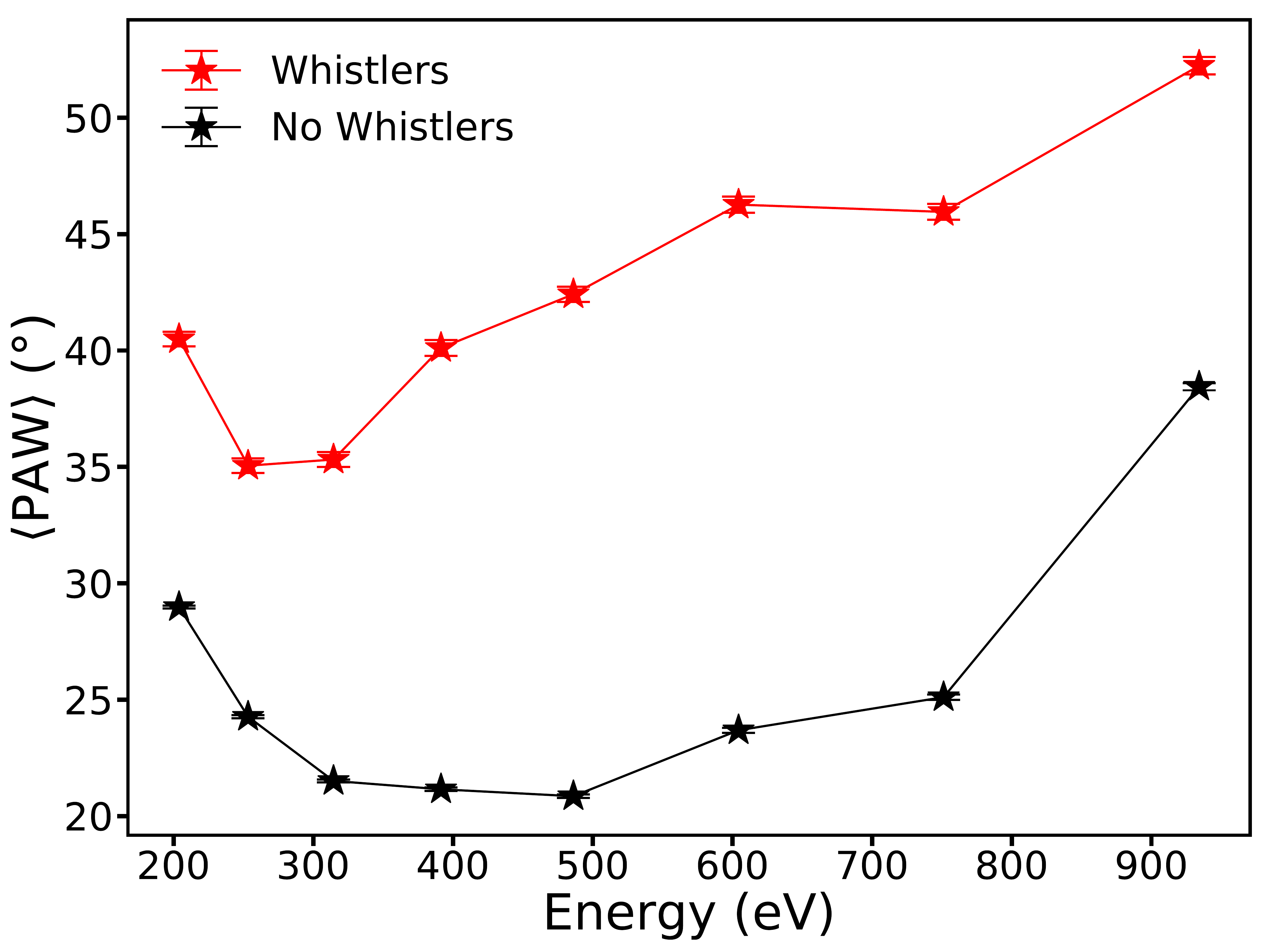}
      \caption{Mean strahl PAW of electrons as a function of electron energy of whistler intervals observed in average BPF data (red) and of non-whistler intervals (black). Error bars show the standard error ($\frac{\sigma}{\sqrt{n}}$).
              }
         \label{fig:PAW_whistelres}
   \end{figure}

In Figure \ref{fig:PAW_whistelres} we show the mean strahl pitch-angle width (PAW) of electrons when the whistlers are observed both in the average BPF data (red) and when the whistlers are not observed at all (black). We observe that the strahl PAWs are significantly larger in all the strahl energy ranges when the whistlers are observed. These observations point toward an interaction between whistler waves and the strahl electrons, which results in the observed broadening of the strahl electron population. The recent study by \citet{Agapitov2020} identified the presence of sunward whistlers along the switchback boundaries which could interact with the anti-sunward strahl electrons and scatter the strahl. However, in our study, we do not have any information on the direction of whistler wave propagation.

\begin{figure}
   \sidecaption
   \includegraphics[width=9cm]{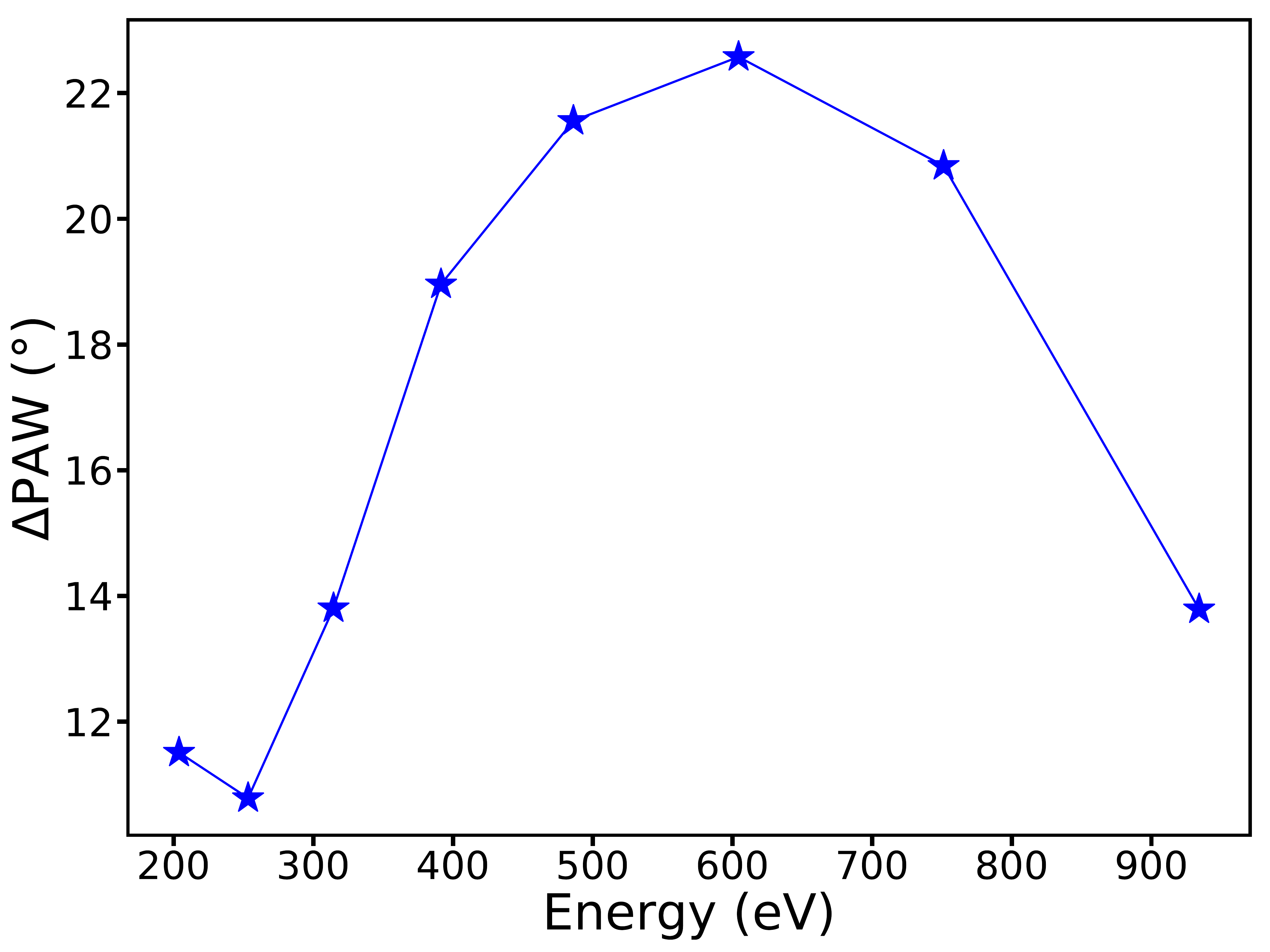}
      \caption{Difference in the mean strahl PAW of electrons of whistlers observed in average BPF data and of non-whistler intervals as a function of electron energy.
              }
         \label{fig:PAW_difference_whistlers}
   \end{figure}

In Figure \ref{fig:PAW_difference_whistlers} we show the difference in the strahl PAW between the intervals with whistlers observed in the average BPF and the intervals with no whistlers. We observe that the PAW is at least 12 degrees broader for sampled energies above 200 eV. The difference between PAW is the largest for the energies between 500 and 700 eV. A similar behavior was observed in the study of \citet{Kajdic2016} at 1 au. An energy-dependent increase in strahl PAW is expected for the resonant interaction with narrowband whistler waves \citep{Behar2020}.

\begin{figure}
   \sidecaption
   \includegraphics[width=9cm]{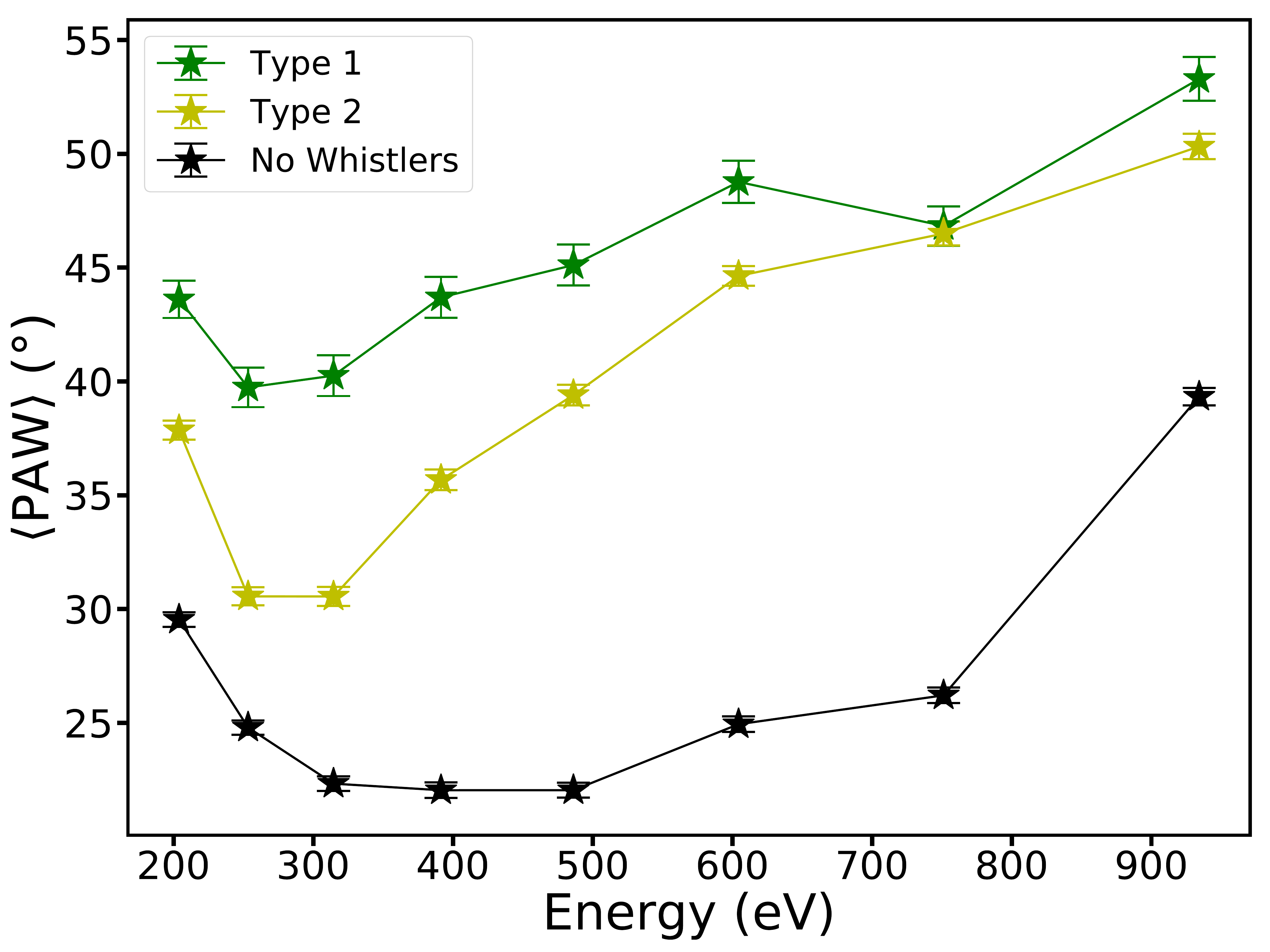}
      \caption{Mean PAW of whistlers separated on the basis of their duration of observation and mean PAW of non-whistlers as a function of electron energy. Type 1 corresponds to the family of whistlers observed consecutively for less than $\sim 3 $s, Type 2 corresponds to the family of whistlers observed consecutively for more than $\sim 20$ s, and the black curve corresponds to the non-whistler intervals. Error bars show the standard error ($\frac{\sigma}{\sqrt{n}}$). 
              }
         \label{fig:PAW_duration_of_whistlers}
   \end{figure}

In Figure \ref{fig:PAW_duration_of_whistlers} we show the mean strahl PAW of electrons corresponding to the whistlers by separating them on the basis of the duration of their consecutive observation. Type 1 corresponds to the family of whistlers which are observed for less than $\sim 3 $ s, and Type 2 corresponds to the family of whistlers which are observed for longer than $\sim 20 $ s. Interestingly, we observe that short duration whistlers show broader PAW than the long duration whistlers. This distinction can be clearly seen in the energy range of 200-600 eV. This is an interesting result as we would expect the long duration whistlers to scatter the strahl broader than the short duration one. Instead, we observe in our study that shorter duration whistlers scatter the strahl more than the ones which are observed for a long duration.

\begin{figure}%
\centering
\subfloat(a){%
\label{fig:first}%
\includegraphics[width=0.8\linewidth]{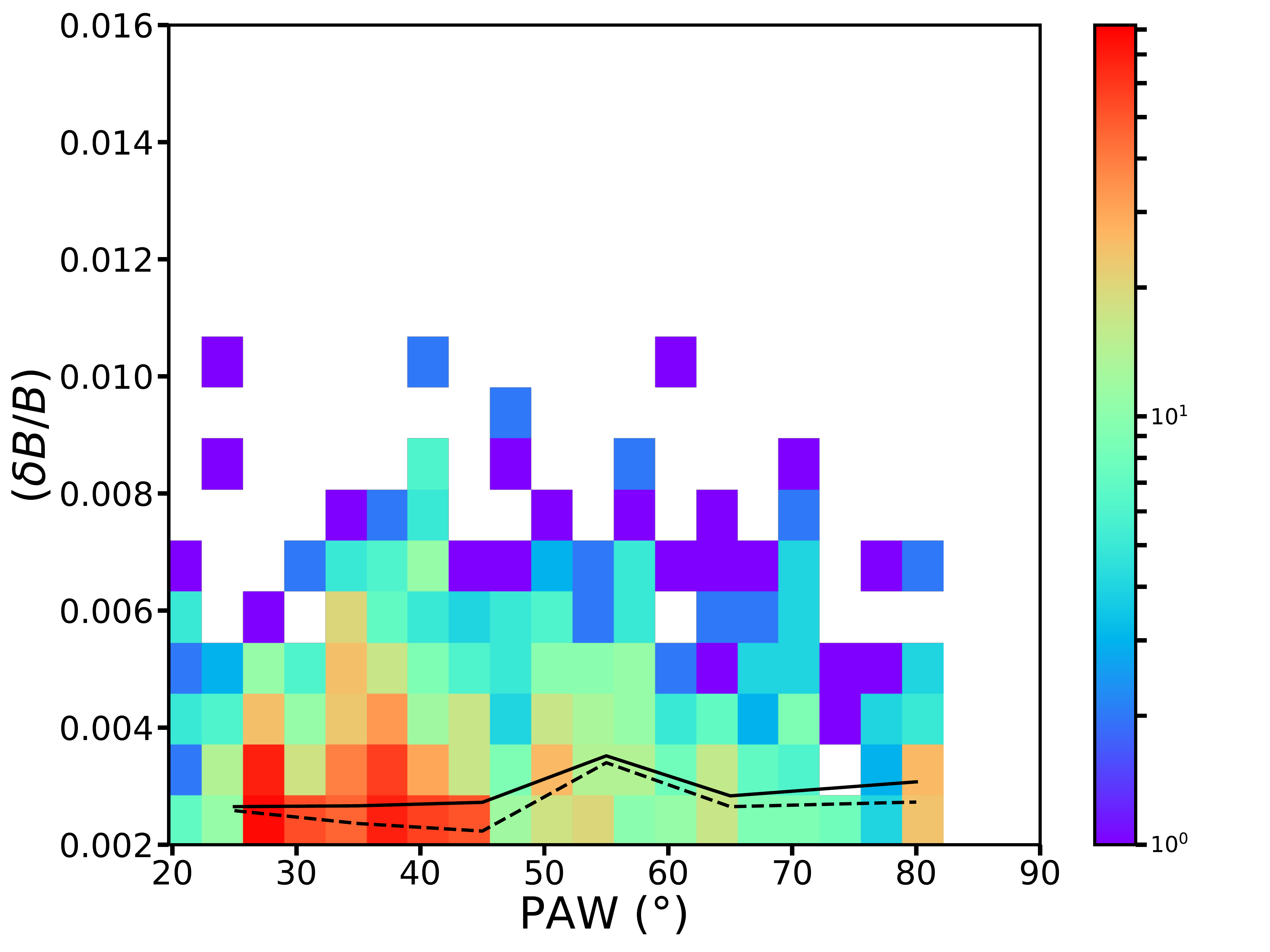}}%
\qquad
\subfloat(b){%
\label{fig:second}%
\includegraphics[width=0.8\linewidth]{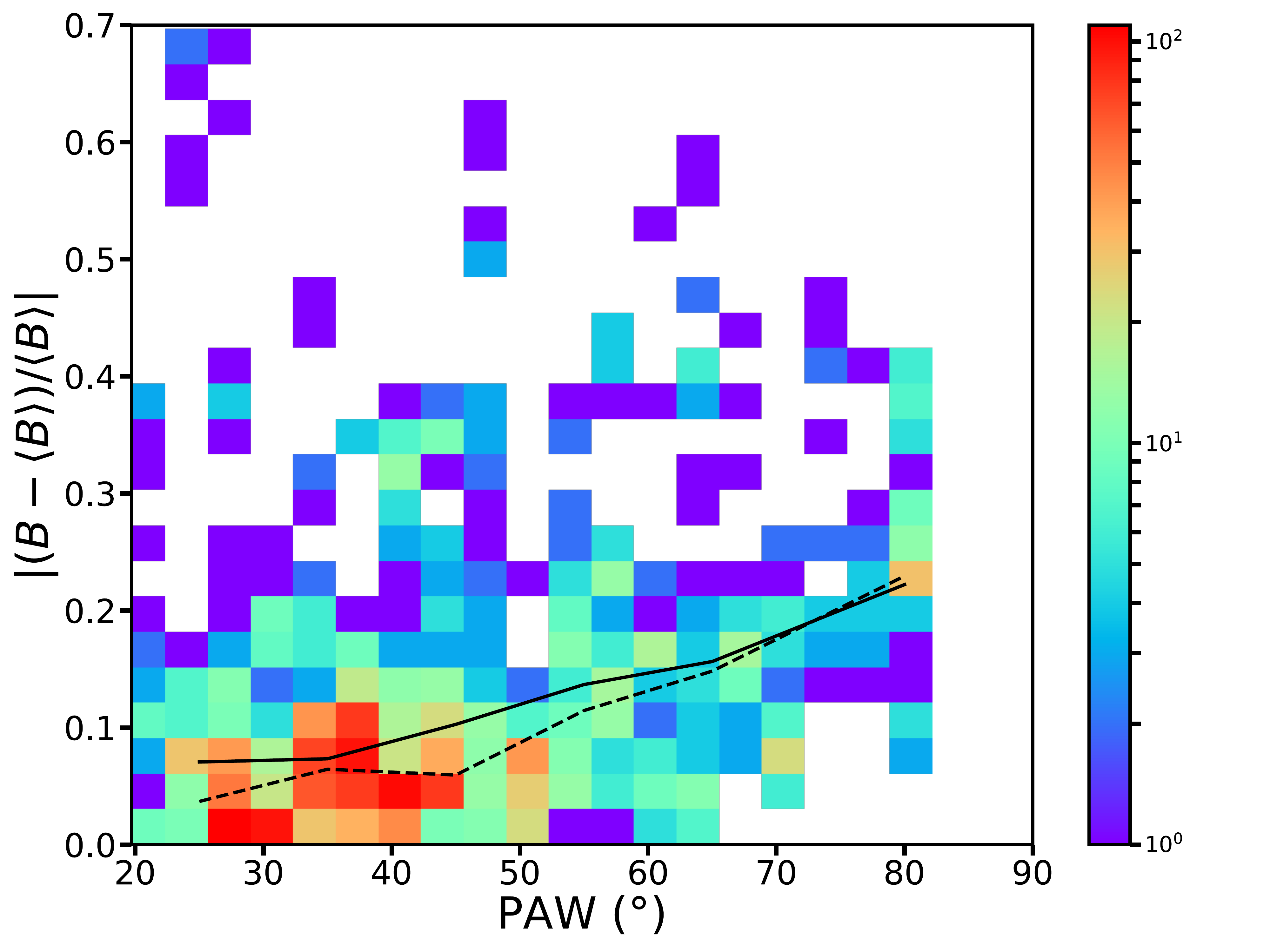}}%
\caption{ 2d histogram of the normalized amplitudes of whistler waves in panel (a) and magnetic field variations corresponding to whistler waves in panel (b) as a function of the strahl PAW of 486 eV electrons. The black line and the dashed line correspond to the mean and median of the whistler wave's normalized amplitudes and the magnetic field variations corresponding to the whistlers, respectively. }
\label{fig:PAW plasma paramteres relation}
\end{figure}

In Figure \ref{fig:PAW plasma paramteres relation} (a), we show the normalized amplitude of whistlers as a function of strahl PAW for electrons of energy 486 eV. We observe that there is no correlation ($\sim 0.06$) between the normalized amplitude of whistlers and strahl PAW. From the trends of normalized amplitudes of whistlers as a function of duration (see Figure \ref{fig:duration_properties} (a)) and no correlation between the normalized amplitude of whistlers and strahl PAW, we can conclude that amplitudes of the waves may not be a reason for the observed differences between the strahl PAW of short duration (Type 1) and long duration (Type 2) whistlers shown in Figure \ref{fig:PAW_duration_of_whistlers}.

In Figure \ref{fig:PAW plasma paramteres relation} (b), we show the normalized magnetic field variations ($\lvert\frac{B-\langle B \rangle}{\langle B \rangle}\rvert$) as a function of strahl PAW for electrons of energy 486 eV. A positive correlation ($\sim 0.54$) between these two parameters was found. We also found a similar correlation between the normalized magnetic field variations and the PAW of strahl electrons for other strahl energies (200 to 700 eV). Interestingly, short duration whistlers have relatively higher normalized magnetic field variations than the long duration ones (see Figure \ref{fig:duration_properties} (d)). These observations suggest that strahl PAW of electrons corresponding to the whistlers that are generated closer to the larger normalized magnetic field variations are broader and that the short duration whistlers are generated close to the larger normalized magnetic field variations, which can be connected to the result in Figure \ref{fig:PAW_duration_of_whistlers}. Therefore, magnetic field variability might be one of the factors for the higher strahl PAW observed for the short duration whistlers compared to the long duration ones as observed in Figure \ref{fig:PAW_duration_of_whistlers}.

The recent study by \citet{Agapitov2020} have shown the presence of oblique whistler waves closer to the Sun, and studies such as \citet{Artemyev2014}, \citet{Artemyev2016}, \citet{Roberg2018}, \citet{Vasko2019}, \citet{Verscharen2019} and \citet{Cattell2020} have suggested that oblique whistlers scatter the strahl electrons better than the parallel ones. Therefore, our observations may suggest that whistlers generated around the relatively higher magnetic field variations might be comparatively more oblique than the ones which are generated around relatively low magnetic field variations.

\begin{figure}
   \sidecaption
   \includegraphics[width=9cm]{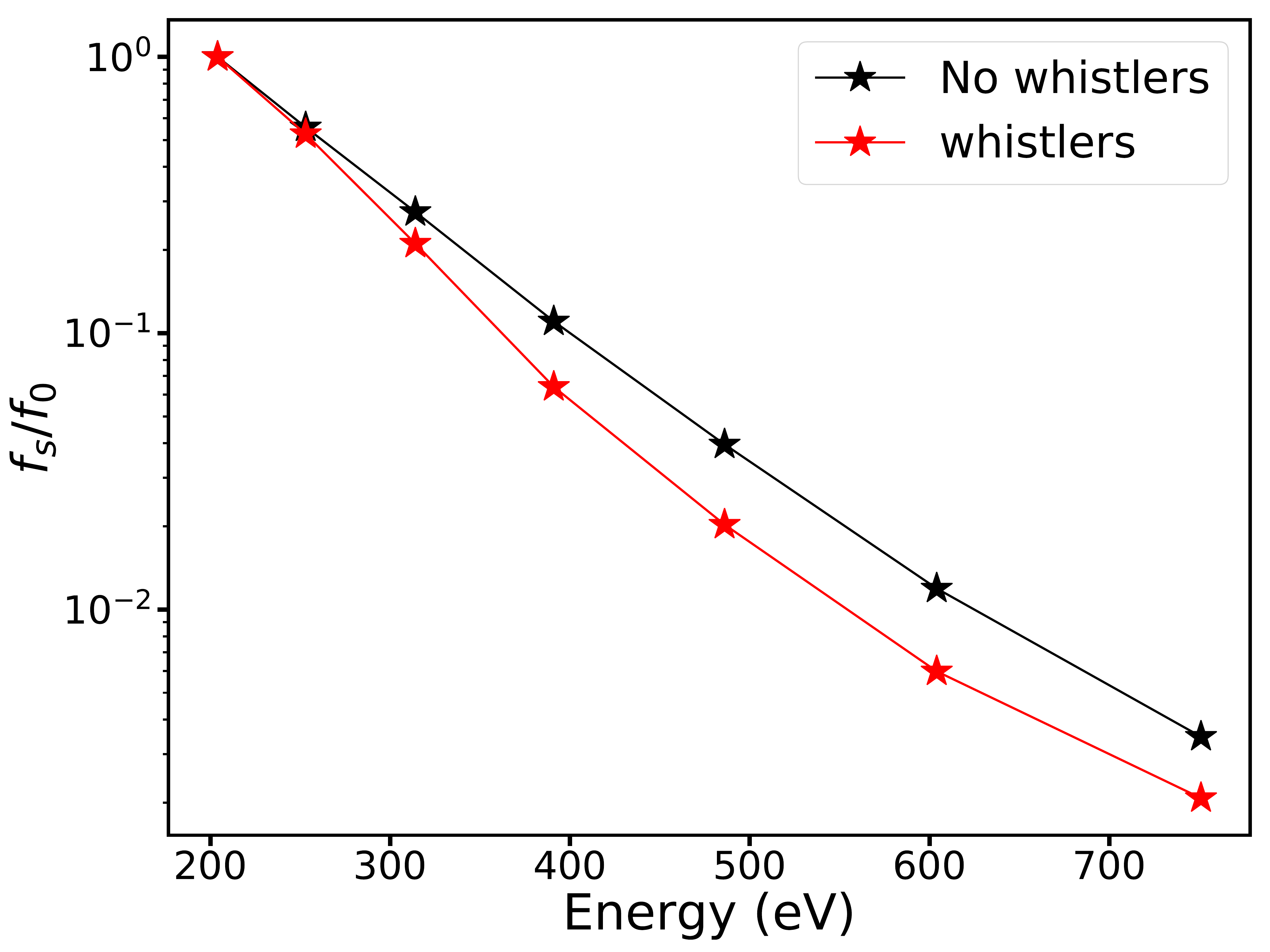}
      \caption{Mean of the parallel cuts through the strahl VDFs ($f_s$) normalized to the VDFs value at 200 eV ($f_0$). The black curve corresponds to the non-whistler intervals, while the red curve corresponds the whistler intervals.  }
         \label{fig:strahl electron distrbution}
   \end{figure}

In Figure \ref{fig:strahl electron distrbution} we show the parallel cut through the strahl electron VDF, that is to say the portion of the strahl velocity distribution function aligned with the magnetic field. We observe that for the non-whistler intervals, the distribution curves can be well represented by a Maxwellian VDF which forms a straight line in parameter space \citep{Halekas2020, Bercic2020}. On the other hand, for the whistler intervals, the distribution is curved compared to the non-whistler cases, corresponding to a Kappa distribution function better ( see Figure 11 in \citet{Bercic2020} for a comparison between Maxwellian and Kappa fits to the strahl parallel VDF). The distribution of strahl electrons was observed to evolve with radial distance; in the near-Sun regions, the strahl was found to be close to a Maxwellian VDF, while further from the Sun it is better represented with a Kappa VDF. Kappa values were found to decrease with radial distance, which means that the relative density of high-energy tails increases as we move away from the Sun. A second suprathermal electron component, the halo, was found to be more important for larger distances from the Sun. The electron halo as well was found to be well represented by a Kappa distribution function \citep{Maksimovic2005,Stverak2009}.
Our observational results reveal that whistler waves can affect the shape of the strahl VDF, and they could be one of the prominent mechanisms responsible for the radial evolution of strahl VDF.

\begin{figure}
   \sidecaption
   \includegraphics[width=9cm]{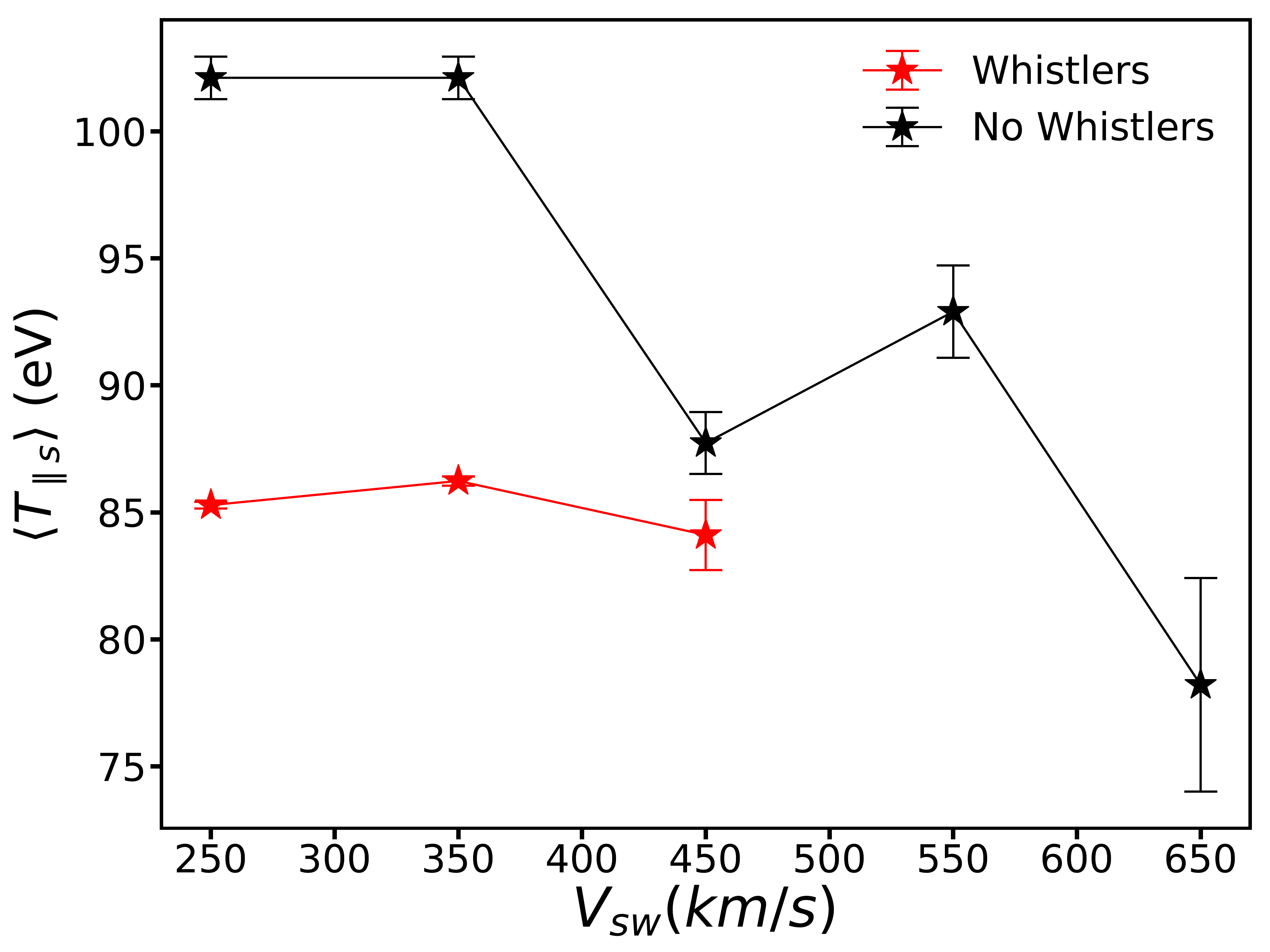}
      \caption{Mean strahl parallel temperatures $(T_{\parallel s})$ as a function of solar wind bulk velocity, for whistlers observed in average BPF data (red) and for non-whistler intervals (black). Error bars show the standard error ($\frac{\sigma}{\sqrt{n}}$).
              }
         \label{fig:whistler strahl temperatures}
   \end{figure}

In Figure \ref{fig:whistler strahl temperatures} we show the mean strahl parallel electron temperatures $(T_{\parallel s})$ corresponding to the whistlers observed in the average BPF data and intervals with no whistlers. We find that mean $T_{\parallel s}$ values are lower for whistler intervals compared to their counterparts of non-whistlers with the same proton bulk velocity. There is more than a $10\%$ decrease in the strahl temperatures corresponding to whistlers when compared to their counterparts corresponding to intervals with no whistlers. As shown by \citet{Bercic2020}, an anticorrelation between $T_{\parallel s}$  and the solar wind velocity can be seen for non-whistler cases. However, this anticorrelation between the $T_{\parallel s}$ and solar wind velocity is not observed for the whistler intervals because these whistlers mainly appear in the slow solar wind.

During the presence of whistler waves, $T_{\parallel s}$ appears to be smaller than at other times. This observation together with the increase in strahl PAW shown in Figure \ref{fig:PAW_whistelres}, leads to the conclusion that during the wave-particle interaction, the parallel strahl electron momentum is converted to perpendicular momentum \citep{Veltri1993}. Further analysis of waves and electron VDFs are required to determine whether or not the total electron energy is conserved during this mechanism.

\section{Conclusions}
Our analysis of PSP DFB BPF data from the first perihelion has shown the presence of bursts of quasi-monochromatic electromagnetic waves. These bursts are observed between 20 and 700 Hz. Despite only one component of the magnetic field data being available and even though the absence of accurate polarization measurements prevents us from accurately characterizing these waves, based on the knowledge from different studies at 1 au and the information from the cross-spectral data analysis, the bursts observed in the PSP's DFB BPF data in the solar wind are interpreted as most likely due to the whistler waves. The statistical study of these wave properties and their relation to the strahl electrons offer a unique opportunity in understanding the significance of the whistlers in strahl electron scattering. These results, in turn, help in gaining the insight into the solar wind energy transport, as strahl electrons carry the majority of the heat flux.

Our study has shown that whistlers occur highly intermittently and the spacecraft central frequencies of the waves are between $f_{LH}$ and $0.2 f_{ce}$. The occurrence probability of whistlers which are observed in the magnetic field is low ($<2 \%$), between 0.17 to 0.26 au whistlers are observed for less than $0.5 \%$ in average BPF data and around $1.5\%$ of the time in peak BPF data. The occurrence of whistlers is highly dependent on the bulk velocity of the solar wind. We observe that the lower the velocity of the solar wind, the higher the occurrence of whistlers. A lower occurrence of whistlers suggests that even though whistlers might play a role in regulating the heat flux, they might not be able to completely explain the regulation of the heat flux in the solar wind.

Around $80\%$ of the whistlers are observed for less than 3 s continuously. The occurrence of long duration whistlers ($> 30$ s) is very low. We show that the velocity of the whistlers is lower for the cases when the whistlers are observed continuously for a long duration. We also show that conditions are found to be quieter, that is to say magnetic field variations such as jumps, drops, and discontinuities are low when the whistler waves are observed continuously for a long duration.

In our study we observe the simultaneous occurrence of whistler and Langmuir waves, which confirms the idea that there might be a common source or a mechanism for the generation of these waves. An in-depth analysis on the reason for the simultaneous presence of whistlers and Langmuir waves should be done in the future to find the common source for both the waves.

The strahl PAW of electrons in the strahl energy ranges are broader when the whistlers are observed, which suggests that whistlers are interacting with the strahl electrons and scattering the strahl. Our observations also show that short duration whistlers scatter the strahl electrons better than the whistlers which are observed for a longer duration. The strahl parallel temperatures are observed to be lower for the intervals corresponding to the whistler waves than to the non-whistler intervals, which suggests that while whistlers are resonantly interacting with the strahl electrons and scattering them, they transfer the momentum from the parallel direction, leading to the decrease in strahl parallel temperatures for whistler intervals. 

Whistler waves were found to have an effect on the shape of the parallel cut through strahl electron VDF. We therefore suggest that whistlers have an important role in the radial evolution of the strahl VDF and the formation of the Kappa-like halo observed farther from the Sun.

\begin{acknowledgements}
Authors thanks M. Liu for helpful discussion. The FIELDS experiment was developed and is operated under NASA contract NNN06AA01C. AL, VK, TD, CF, VKJ and CR acknowledge financial support of CNES in the frame of Parker Solar Probe grant. O.A. was supported by NASA grants 80NNSC19K0848, 80NSSC20K0697, 80NSSC20K0218. SDB acknowledges the support of the Leverhulme Trust Visiting Professorship programme. Parker Solar Probe was designed, built, and is now operated by the Johns Hopkins Applied Physics Laboratory as part of NASA’s Living with a Star (LWS) program (contract NNN06AA01C). Support from the LWS management and technical team has played a critical role in the success of the Parker Solar Probe mission. The data used in this study are available at the NASA Space Physics Data Facility (SPDF), \url{https://spdf.gsfc.nasa.gov}. Figures were
produced using Matplotlib v3.1.3 \citep{Hunter2007,caswell2020}
\end{acknowledgements}

\bibliographystyle{aa.bst} 
\bibliography{References}

\begin{thebibliography}{55}
\expandafter\ifx\csname natexlab\endcsname\relax\def\natexlab#1{#1}\fi

\bibitem[{{Agapitov} {et~al.}(2020){Agapitov}, {Dudok de Wit}, {Mozer},
  {Bonnell}, {Drake}, {Malaspina}, {Krasnoselskikh}, {Bale}, {Whittlesey},
  {Case}, {Chaston}, {Froment}, {Goetz}, {Goodrich}, {Harvey}, {Kasper},
  {Korreck}, {Larson}, {Livi}, {MacDowall}, {Pulupa}, {Revillet}, {Stevens}, \&
  {Wygant}}]{Agapitov2020}
{Agapitov}, O.~V., {Dudok de Wit}, T., {Mozer}, F.~S., {et~al.} 2020, \apjl,
  891, L20

\bibitem[{{Alexandrova} {et~al.}(2020){Alexandrova}, {Krishna Jagarlamudi},
  {Rossi}, {Maksimovic}, {Hellinger}, {Shprits}, \&
  {Mangeney}}]{Alexandrova2020}
{Alexandrova}, O., {Krishna Jagarlamudi}, V., {Rossi}, C., {et~al.} 2020, arXiv
  e-prints, arXiv:2004.01102

\bibitem[{{Alexandrova} {et~al.}(2012){Alexandrova}, {Lacombe}, {Mangeney},
  {Grappin}, \& {Maksimovic}}]{Alexandrova2012}
{Alexandrova}, O., {Lacombe}, C., {Mangeney}, A., {Grappin}, R., \&
  {Maksimovic}, M. 2012, The Astrophysical Journal, 760, 121

\bibitem[{{Artemyev} {et~al.}(2016){Artemyev}, {Agapitov}, {Mourenas},
  {Krasnoselskikh}, {Shastun}, \& {Mozer}}]{Artemyev2016}
{Artemyev}, A., {Agapitov}, O., {Mourenas}, D., {et~al.} 2016, \ssr, 200, 261

\bibitem[{{Artemyev} {et~al.}(2014){Artemyev}, {Vasiliev}, {Mourenas},
  {Agapitov}, \& {Krasnoselskikh}}]{Artemyev2014}
{Artemyev}, A.~V., {Vasiliev}, A.~A., {Mourenas}, D., {Agapitov}, O.~V., \&
  {Krasnoselskikh}, V.~V. 2014, Physics of Plasmas, 21, 102903

\bibitem[{{Bale} {et~al.}(2019){Bale}, {Badman}, {Bonnell}, {Bowen}, {Burgess},
  {Case}, {Cattell}, {Chandran}, {Chaston}, {Chen}, {Drake}, {de Wit},
  {Eastwood}, {Ergun}, {Farrell}, {Fong}, {Goetz}, {Goldstein}, {Goodrich},
  {Harvey}, {Horbury}, {Howes}, {Kasper}, {Kellogg}, {Klimchuk}, {Korreck},
  {Krasnoselskikh}, {Krucker}, {Laker}, {Larson}, {MacDowall}, {Maksimovic},
  {Malaspina}, {Martinez-Oliveros}, {McComas}, {Meyer-Vernet}, {Moncuquet},
  {Mozer}, {Phan}, {Pulupa}, {Raouafi}, {Salem}, {Stansby}, {Stevens}, {Szabo},
  {Velli}, {Woolley}, \& {Wygant}}]{bale19}
{Bale}, S.~D., {Badman}, S.~T., {Bonnell}, J.~W., {et~al.} 2019, \nat, 576, 237

\bibitem[{{Bale} {et~al.}(2016){Bale}, {Goetz}, {Harvey}, {Turin}, {Bonnell},
  {Dudok de Wit}, {Ergun}, {MacDowall}, {Pulupa}, {Andre}, {Bolton},
  {Bougeret}, {Bowen}, {Burgess}, {Cattell}, {Chandran}, {Chaston}, {Chen},
  {Choi}, {Connerney}, {Cranmer}, {Diaz-Aguado}, {Donakowski}, {Drake},
  {Farrell}, {Fergeau}, {Fermin}, {Fischer}, {Fox}, {Glaser}, {Goldstein},
  {Gordon}, {Hanson}, {Harris}, {Hayes}, {Hinze}, {Hollweg}, {Horbury},
  {Howard}, {Hoxie}, {Jannet}, {Karlsson}, {Kasper}, {Kellogg}, {Kien},
  {Klimchuk}, {Krasnoselskikh}, {Krucker}, {Lynch}, {Maksimovic}, {Malaspina},
  {Marker}, {Martin}, {Martinez-Oliveros}, {McCauley}, {McComas}, {McDonald},
  {Meyer-Vernet}, {Moncuquet}, {Monson}, {Mozer}, {Murphy}, {Odom},
  {Oliverson}, {Olson}, {Parker}, {Pankow}, {Phan}, {Quataert}, {Quinn},
  {Ruplin}, {Salem}, {Seitz}, {Sheppard}, {Siy}, {Stevens}, {Summers}, {Szabo},
  {Timofeeva}, {Vaivads}, {Velli}, {Yehle}, {Werthimer}, \&
  {Wygant}}]{Bale2016}
{Bale}, S.~D., {Goetz}, K., {Harvey}, P.~R., {et~al.} 2016, \ssr, 204, 49

\bibitem[{{Behar} {et~al.}(2020){Behar}, {Sahraoui}, \& {Bercic}}]{Behar2020}
{Behar}, E., {Sahraoui}, F., \& {Bercic}, L. 2020, arXiv e-prints,
  arXiv:2004.09130

\bibitem[{{Ber{\v{c}}i{\v{c}}} {et~al.}(2020){Ber{\v{c}}i{\v{c}}}, {Larson},
  {Whittlesey}, {Maksimovi{\'c}}, {Badman}, {Landi}, {Matteini}, {Bale},
  {Bonnell}, {Case}, {Dudok de Wit}, {Goetz}, {Harvey}, {Kasper}, {Korreck},
  {Livi}, {MacDowall}, {Malaspina}, {Pulupa}, \& {Stevens}}]{Bercic2020}
{Ber{\v{c}}i{\v{c}}}, L., {Larson}, D., {Whittlesey}, P., {et~al.} 2020, \apj,
  892, 88

\bibitem[{{Ber{\v{c}}i{\v{c}}} {et~al.}(2019){Ber{\v{c}}i{\v{c}}},
  {Maksimovi{\'c}}, {}, {Land i}, \& {Matteini}}]{Laura2019}
{Ber{\v{c}}i{\v{c}}}, L., {Maksimovi{\'c}}, {}, M., {Land i}, S., \&
  {Matteini}, L. 2019, \mnras, 486, 3404

\bibitem[{{Boldyrev} \& {Horaites}(2019)}]{Boldyrev2019}
{Boldyrev}, S. \& {Horaites}, K. 2019, \mnras, 489, 3412

\bibitem[{Case {et~al.}(2020)Case, Kasper, Stevens, Korreck, Paulson, Daigneau,
  Caldwell, Freeman, Henry, Klingensmith, Bookbinder, Robinson, Berg, Tiu,
  Wright, Reinhart, Curtis, Ludlam, Larson, Whittlesey, Livi, Klein, \&
  Martinović}]{case_solar_2020}
Case, A.~W., Kasper, J.~C., Stevens, M.~L., {et~al.} 2020, The Astrophysical
  Journal Supplement Series, 246, 43

\bibitem[{Caswell {et~al.}(2020)Caswell, Droettboom, Lee, Hunter, Firing,
  Stansby, Klymak, Hoffmann, de~Andrade, Varoquaux, Nielsen, Root, Elson, May,
  Dale, Lee, Seppänen, McDougall, Straw, Hobson, Gohlke, Yu, Ma, Vincent,
  Silvester, Moad, Kniazev, Ivanov, Ernest, \& Katins}]{caswell2020}
Caswell, T.~A., Droettboom, M., Lee, A., {et~al.} 2020, matplotlib/matplotlib
  v3.1.3

\bibitem[{{Cattell} {et~al.}(2020){Cattell}, {Short}, {Breneman}, \&
  {Grul}}]{Cattell2020}
{Cattell}, C.~A., {Short}, B., {Breneman}, A.~W., \& {Grul}, P. 2020, \apj,
  897, 126

\bibitem[{{Dudok de Wit} {et~al.}(2020){Dudok de Wit}, {Krasnoselskikh},
  {Bale}, {Bonnell}, {Bowen}, {Chen}, {Froment}, {Goetz}, {Harvey},
  {Jagarlamudi}, {Larosa}, {MacDowall}, {Malaspina}, {Matthaeus}, {Pulupa},
  {Velli}, \& {Whittlesey}}]{DudokdeWit2020}
{Dudok de Wit}, T., {Krasnoselskikh}, V.~V., {Bale}, S.~D., {et~al.} 2020,
  \apjs, 246, 39

\bibitem[{{Feldman} {et~al.}(1978){Feldman}, {Asbridge}, {Bame}, {Gosling}, \&
  {Lemons}}]{Feldman1978}
{Feldman}, W.~C., {Asbridge}, J.~R., {Bame}, S.~J., {Gosling}, J.~T., \&
  {Lemons}, D.~S. 1978, \jgr, 83, 5285

\bibitem[{Fox {et~al.}(2016)Fox, Velli, Bale, Decker, Driesman, Howard, Kasper,
  Kinnison, Kusterer, Lario, Lockwood, McComas, Raouafi, \&
  Szabo}]{fox_solar_2016}
Fox, N.~J., Velli, M.~C., Bale, S.~D., {et~al.} 2016, Space Science Reviews,
  204, 7

\bibitem[{Froment {et~al.}(2020)Froment, Krasnoselskikh, Agapitov, de~Wit,
  Malaspina, Jagarlamudi, Kretzschmar, Larosa, Bale, Bonnell,
  {et~al.}}]{froment2020whistler}
Froment, C., Krasnoselskikh, V., Agapitov, O.~V., {et~al.} 2020, in AGU Fall
  Meeting 2020, AGU

\bibitem[{{Gary} {et~al.}(1994){Gary}, {Scime}, {Phillips}, \&
  {Feldman}}]{Gary1994JGR}
{Gary}, S.~P., {Scime}, E.~E., {Phillips}, J.~L., \& {Feldman}, W.~C. 1994,
  \jgr, 99, 23391

\bibitem[{{Graham} {et~al.}(2017){Graham}, {Rae}, {Owen}, {Walsh}, {Arridge},
  {Gilbert}, {Lewis}, {Jones}, {Forsyth}, {Coates}, \& {Waite}}]{Graham2017}
{Graham}, G.~A., {Rae}, I.~J., {Owen}, C.~J., {et~al.} 2017, Journal of
  Geophysical Research (Space Physics), 122, 3858

\bibitem[{{Halekas} {et~al.}(2020{\natexlab{a}}){Halekas}, {Whittlesey},
  {Larson}, {McGinnis}, {Maksimovic}, {Berthomier}, {Kasper}, {Case},
  {Korreck}, {Stevens}, {Klein}, {Bale}, {MacDowall}, {Pulupa}, {Malaspina},
  {Goetz}, \& {Harvey}}]{Halekas2020}
{Halekas}, J.~S., {Whittlesey}, P., {Larson}, D.~E., {et~al.}
  2020{\natexlab{a}}, \apjs, 246, 22

\bibitem[{{Halekas} {et~al.}(2020{\natexlab{b}}){Halekas}, {Whittlesey},
  {Larson}, {McGinnis}, {Bale}, {Berthomier}, {Case}, {Chandran}, {Kasper},
  {Klein}, {Korreck}, {Livi}, {MacDowall}, {Maksimovic}, {Malaspina},
  {Matteini}, {Pulupa}, \& {Stevens}}]{Halekas2020arXiv}
{Halekas}, J.~S., {Whittlesey}, P.~L., {Larson}, D.~E., {et~al.}
  2020{\natexlab{b}}, arXiv e-prints, arXiv:2010.10302

\bibitem[{{Hammond} {et~al.}(1996){Hammond}, {Feldman}, {McComas}, {Phillips},
  \& {Forsyth}}]{Hammond1996}
{Hammond}, C.~M., {Feldman}, W.~C., {McComas}, D.~J., {Phillips}, J.~L., \&
  {Forsyth}, R.~J. 1996, Astronomy and Astrophysics, 316, 350

\bibitem[{Hunter(2007)}]{Hunter2007}
Hunter, J.~D. 2007, Computing in Science \& Engineering, 9, 90

\bibitem[{{Jagarlamudi} {et~al.}(2020){Jagarlamudi}, {Alexandrova},
  {Ber{\v{c}}i{\v{c}}}, {de Wit}, {Krasnoselskikh}, {Maksimovic}, \&
  {{\v{S}}tver{\'a}k}}]{Jagarlamudi2020}
{Jagarlamudi}, V.~K., {Alexandrova}, O., {Ber{\v{c}}i{\v{c}}}, L., {et~al.}
  2020, \apj, 897, 118

\bibitem[{Jannet {et~al.}(2020)Jannet, Dudok~de Wit, Krasnoselskikh,
  Kretzschmar, Fergeau, Bergerard-Timofeeva, Agrapart, Brochot, Chalumeau,
  Martin, Revillet, Bale, Maksimovic, Bowen, Brysbaert, Goetz, Guilhem, Harvey,
  Leray, \& Lorfevre}]{jannet2020}
Jannet, G., Dudok~de Wit, T., Krasnoselskikh, V., {et~al.} 2020, Journal of
  Geophysical Research: Space Physics, n/a, e2020JA028543, e2020JA028543
  2020JA028543

\bibitem[{{Kajdi{\v c}} {et~al.}(2016){Kajdi{\v c}}, {Alexandrova},
  {Maksimovic}, {Lacombe}, \& {Fazakerley}}]{Kajdic2016}
{Kajdi{\v c}}, P., {Alexandrova}, O., {Maksimovic}, M., {Lacombe}, C., \&
  {Fazakerley}, A.~N. 2016, The Astrophysical Journal, 833, 172

\bibitem[{{Kasper} {et~al.}(2016){Kasper}, {Abiad}, {Austin}, {Balat-Pichelin},
  {Bale}, {Belcher}, {Berg}, {Bergner}, {Berthomier}, {Bookbinder}, {Brodu},
  {Caldwell}, {Case}, {Chand ran}, {Cheimets}, {Cirtain}, {Cranmer}, {Curtis},
  {Daigneau}, {Dalton}, {Dasgupta}, {DeTomaso}, {Diaz-Aguado}, {Djordjevic},
  {Donaskowski}, {Effinger}, {Florinski}, {Fox}, {Freeman}, {Gallagher},
  {Gary}, {Gauron}, {Gates}, {Goldstein}, {Golub}, {Gordon}, {Gurnee}, {Guth},
  {Halekas}, {Hatch}, {Heerikuisen}, {Ho}, {Hu}, {Johnson}, {Jordan},
  {Korreck}, {Larson}, {Lazarus}, {Li}, {Livi}, {Ludlam}, {Maksimovic},
  {McFadden}, {Marchant}, {Maruca}, {McComas}, {Messina}, {Mercer}, {Park},
  {Peddie}, {Pogorelov}, {Reinhart}, {Richardson}, {Robinson}, {Rosen},
  {Skoug}, {Slagle}, {Steinberg}, {Stevens}, {Szabo}, {Taylor}, {Tiu}, {Turin},
  {Velli}, {Webb}, {Whittlesey}, {Wright}, {Wu}, \& {Zank}}]{Kasper2016}
{Kasper}, J.~C., {Abiad}, R., {Austin}, G., {et~al.} 2016, \ssr, 204, 131

\bibitem[{{Kasper} {et~al.}(2019){Kasper}, {Bale}, {Belcher}, {Berthomier},
  {Case}, {Chandran}, {Curtis}, {Gallagher}, {Gary}, {Golub}, {Halekas}, {Ho},
  {Horbury}, {Hu}, {Huang}, {Klein}, {Korreck}, {Larson}, {Livi}, {Maruca},
  {Lavraud}, {Louarn}, {Maksimovic}, {Martinovic}, {McGinnis}, {Pogorelov},
  {Richardson}, {Skoug}, {Steinberg}, {Stevens}, {Szabo}, {Velli},
  {Whittlesey}, {Wright}, {Zank}, {MacDowall}, {McComas}, {McNutt}, {Pulupa},
  {Raouafi}, \& {Schwadron}}]{kasper19}
{Kasper}, J.~C., {Bale}, S.~D., {Belcher}, J.~W., {et~al.} 2019, \nat, 576, 228

\bibitem[{{Kennel} {et~al.}(1980){Kennel}, {Scarf}, {Coroniti}, {Fredricks},
  {Gurnett}, \& {Smith}}]{Kennel1980}
{Kennel}, C.~F., {Scarf}, F.~L., {Coroniti}, F.~V., {et~al.} 1980, \grl, 7, 129

\bibitem[{{Lacombe} {et~al.}(2014){Lacombe}, {Alexandrova}, {Matteini},
  {Santol{\'{\i}}k}, {Cornilleau-Wehrlin}, {Mangeney}, {de Conchy}, \&
  {Maksimovic}}]{Lacombe2014}
{Lacombe}, C., {Alexandrova}, O., {Matteini}, L., {et~al.} 2014, The
  Astrophysical Journal, 796, 5

\bibitem[{Larosa {et~al.}(2020)Larosa, Krasnoselskikh, Dudok~de Wit, Agapitov,
  Froment, Jagarlamudi, Velli, Bale, Case, Goetz, Harvey, Kasper, Korreck,
  Larson, MacDowall, Malaspina, Pulupa, Revillet, \&
  Stevens}]{larosa_switchbacks_2020}
Larosa, A., Krasnoselskikh, V., Dudok~de Wit, T., {et~al.} 2020, under review,
  Astronomy and Astrophysics, this issue, arXiv:2012.10420

\bibitem[{{Maksimovic} {et~al.}(2005){Maksimovic}, {Zouganelis}, {Chaufray},
  {Issautier}, {Scime}, {Littleton}, {Marsch}, {McComas}, {Salem}, {Lin}, \&
  {Elliott}}]{Maksimovic2005}
{Maksimovic}, M., {Zouganelis}, I., {Chaufray}, J.-Y., {et~al.} 2005, Journal
  of Geophysical Research (Space Physics), 110, A09104

\bibitem[{{Malaspina} {et~al.}(2016){Malaspina}, {Ergun}, {Bolton}, {Kien},
  {Summers}, {Stevens}, {Yehle}, {Karlsson}, {Hoxie}, {Bale}, \&
  {Goetz}}]{Malaspina2016}
{Malaspina}, D.~M., {Ergun}, R.~E., {Bolton}, M., {et~al.} 2016, Journal of
  Geophysical Research (Space Physics), 121, 5088

\bibitem[{{Moncuquet} {et~al.}(2020){Moncuquet}, {Meyer-Vernet}, {Issautier},
  {Pulupa}, {Bonnell}, {Bale}, {de Wit}, {Goetz}, {Griton}, {Harvey},
  {MacDowall}, {Maksimovic}, \& {Malaspina}}]{moncuquet20}
{Moncuquet}, M., {Meyer-Vernet}, N., {Issautier}, K., {et~al.} 2020, \apjs,
  246, 44

\bibitem[{{Ogilvie} \& {Scudder}(1978)}]{Ogilvie1978JGR}
{Ogilvie}, K.~W. \& {Scudder}, J.~D. 1978, \jgr, 83, 3776

\bibitem[{{Pagel} {et~al.}(2007){Pagel}, {Gary}, {de Koning}, {Skoug}, \&
  {Steinberg}}]{Pagel2007}
{Pagel}, C., {Gary}, S.~P., {de Koning}, C.~A., {Skoug}, R.~M., \& {Steinberg},
  J.~T. 2007, Journal of Geophysical Research (Space Physics), 112, A04103

\bibitem[{{Pilipp} {et~al.}(1987{\natexlab{a}}){Pilipp}, {Miggenrieder},
  {Montgomery}, {M{\"u}hlh{\"a}user}, {Rosenbauer}, \& {Schwenn}}]{Pilipp1987b}
{Pilipp}, W.~G., {Miggenrieder}, H., {Montgomery}, M.~D., {et~al.}
  1987{\natexlab{a}}, \jgr, 92, 1075

\bibitem[{{Pilipp} {et~al.}(1987{\natexlab{b}}){Pilipp}, {Miggenrieder},
  {M{\"u}hlha{\"u}ser}, {Rosenbauer}, {Schwenn}, \& {Neubauer}}]{Pillip1987a}
{Pilipp}, W.~G., {Miggenrieder}, H., {M{\"u}hlha{\"u}ser}, K.~H., {et~al.}
  1987{\natexlab{b}}, \jgr, 92, 1103

\bibitem[{{Pulupa} {et~al.}(2017){Pulupa}, {Bale}, {Bonnell}, {Bowen},
  {Carruth}, {Goetz}, {Gordon}, {Harvey}, {Maksimovic},
  {Mart{\'\i}nez-Oliveros}, {Moncuquet}, {Saint-Hilaire}, {Seitz}, \&
  {Sundkvist}}]{Pulupa2017}
{Pulupa}, M., {Bale}, S.~D., {Bonnell}, J.~W., {et~al.} 2017, Journal of
  Geophysical Research (Space Physics), 122, 2836

\bibitem[{{Roberg-Clark} {et~al.}(2018){Roberg-Clark}, {Drake}, {Swisdak}, \&
  {Reynolds}}]{Roberg2018}
{Roberg-Clark}, G.~T., {Drake}, J.~F., {Swisdak}, M., \& {Reynolds}, C.~S.
  2018, The Astrophysical Journal, 867, 154

\bibitem[{{Scime} {et~al.}(1994){Scime}, {Bame}, {Feldman}, {Gary}, {Phillips},
  \& {Balogh}}]{Scime1994JGR}
{Scime}, E.~E., {Bame}, S.~J., {Feldman}, W.~C., {et~al.} 1994, \jgr, 99, 23401

\bibitem[{{Stansby} {et~al.}(2016){Stansby}, {Horbury}, {Chen}, \&
  {Matteini}}]{Stansby2016}
{Stansby}, D., {Horbury}, T.~S., {Chen}, C.~H.~K., \& {Matteini}, L. 2016, The
  Astrophysical Journal Letters, 829, L16

\bibitem[{{Tang} {et~al.}(2020){Tang}, {Zank}, \& {Kolobov}}]{Tang2020}
{Tang}, B., {Zank}, G.~P., \& {Kolobov}, V.~I. 2020, \apj, 892, 95

\bibitem[{{Tong} {et~al.}(2019{\natexlab{a}}){Tong}, {Vasko}, {Artemyev},
  {Bale}, \& {Mozer}}]{Tong2019stasticalstudy}
{Tong}, Y., {Vasko}, I.~Y., {Artemyev}, A.~V., {Bale}, S.~D., \& {Mozer}, F.~S.
  2019{\natexlab{a}}, The Astrophysical Journal, 878, 41

\bibitem[{{Tong} {et~al.}(2019{\natexlab{b}}){Tong}, {Vasko}, {Pulupa},
  {Mozer}, {Bale}, {Artemyev}, \& {Krasnoselskikh}}]{Tong2019}
{Tong}, Y., {Vasko}, I.~Y., {Pulupa}, M., {et~al.} 2019{\natexlab{b}}, The
  Astrophysical Journal Letters, 870, L6

\bibitem[{{{\v S}tver{\'a}k} {et~al.}(2009){{\v S}tver{\'a}k}, {Maksimovic},
  {Tr{\'a}vn{\'{\i}}{\v c}ek}, {Marsch}, {Fazakerley}, \&
  {Scime}}]{Stverak2009}
{{\v S}tver{\'a}k}, {\v S}., {Maksimovic}, M., {Tr{\'a}vn{\'{\i}}{\v c}ek},
  P.~M., {et~al.} 2009, Journal of Geophysical Research (Space Physics), 114,
  A05104

\bibitem[{{{\v S}tver{\'a}k} {et~al.}(2015){{\v S}tver{\'a}k},
  {Tr{\'a}vn{\'{\i}}{\v c}ek}, \& {Hellinger}}]{Stverak2015}
{{\v S}tver{\'a}k}, {\v S}., {Tr{\'a}vn{\'{\i}}{\v c}ek}, P.~M., \&
  {Hellinger}, P. 2015, Journal of Geophysical Research (Space Physics), 120,
  8177

\bibitem[{{Vasko} {et~al.}(2019){Vasko}, {Krasnoselskikh}, {Tong}, {Bale},
  {Bonnell}, \& {Mozer}}]{Vasko2019}
{Vasko}, I.~Y., {Krasnoselskikh}, V., {Tong}, Y., {et~al.} 2019, \apjl, 871,
  L29

\bibitem[{{Veltri} \& {Zimbardo}(1993)}]{Veltri1993}
{Veltri}, P. \& {Zimbardo}, G. 1993, Journal of Geophysical Research (Space
  Physics), 98, 13325

\bibitem[{{Verscharen} {et~al.}(2019){Verscharen}, {Chandran}, {Jeong},
  {Salem}, {Pulupa}, \& {Bale}}]{Verscharen2019}
{Verscharen}, D., {Chandran}, B. D.~G., {Jeong}, S.-Y., {et~al.} 2019, \apj,
  886, 136

\bibitem[{{Vocks}(2012)}]{Vocks2012}
{Vocks}, C. 2012, Space Science Reviews, 172, 303

\bibitem[{{Vocks} \& {Mann}(2003)}]{Vocks2003}
{Vocks}, C. \& {Mann}, G. 2003, The Astrophysical Journal, 593, 1134

\bibitem[{Whittlesey {et~al.}(2020)Whittlesey, Larson, Kasper, Halekas,
  Abatcha, Abiad, Berthomier, Case, Chen, Curtis, Dalton, Klein, Korreck, Livi,
  Ludlam, Marckwordt, Rahmati, Robinson, Slagle, Stevens, Tiu, \&
  Verniero}]{whittlesey_solar_2020}
Whittlesey, P.~L., Larson, D.~E., Kasper, J.~C., {et~al.} 2020, The
  Astrophysical Journal Supplement Series, 246, 74

\bibitem[{{Zhang} {et~al.}(1998){Zhang}, {Matsumoto}, \& {Kojima}}]{Zhang1998}
{Zhang}, Y., {Matsumoto}, H., \& {Kojima}, H. 1998, Journal of Geophysical
  Research (Space Physics), 103, 20529

\end{thebibliography}

\end{document}